\documentclass[aps,preprint,nofootinbib,eqsecnum,prb]{revtex4}
\usepackage{amsmath,amssymb,bm}
\usepackage{graphicx}
\usepackage{epstopdf}
\usepackage{latexsym}
\usepackage{setspace}   

\newcommand{\beq}{\begin{equation}}
\newcommand{\eeq}{\end{equation}}
\newcommand{\beqn}{\begin{eqnarray}}
\newcommand{\eeqn}{\end{eqnarray}}

\begin{document}

\title{Global phase diagrams of frustrated quantum antiferromagnets in two dimensions:
doubled Chern-Simons theory}
\author{Cenke Xu}
\affiliation{Department of Physics, Harvard University, Cambridge MA 02138, USA}
\author{Subir Sachdev}
\affiliation{Department of Physics, Harvard University, Cambridge MA 02138, USA}
\date{\today\\
\vspace{1.6in}}

\begin{abstract}
~\\
We present a general approach to understanding the quantum phases
and phase transitions of quantum antiferromagnets in two spatial
dimensions. We begin with the simplest spin liquid state, the
$Z_2$ spin liquid, whose elementary excitations are spinons and
visons, carrying $Z_2$ electric and magnetic charges respectively.
Their dynamics are expressed in terms of a doubled U(1) Chern
Simons theory, which correctly captures the `topological' order of
the $Z_2$ spin liquid state. We show that the same theory also
yields a description of the variety of ordered phases obtained
when one or more of the elementary excitations condense. Field
theories for the transitions and multicritical points between
these phases are obtained. We survey experimental results on
antiferromagnets on the anisotropic triangular lattice, and make
connections between their phase diagrams and our results.
~\\
\end{abstract}

\maketitle

\section{Introduction}
\label{sec:intro}

The study of exotic phases of quantum antiferromagnets has
received a great impetus by the experimental discovery of a number
of candidate $S=1/2$ Mott insulators. The primary aim of our paper
is to present an attempt to place the experimentally discovered
phases in a single global phase diagram; such a phase diagram
exposes new relations between the excitations of the various
phases, and leads to theories for the possible quantum phase
transitions between them.

\begin{figure}
\begin{center}
\includegraphics[width=4in]{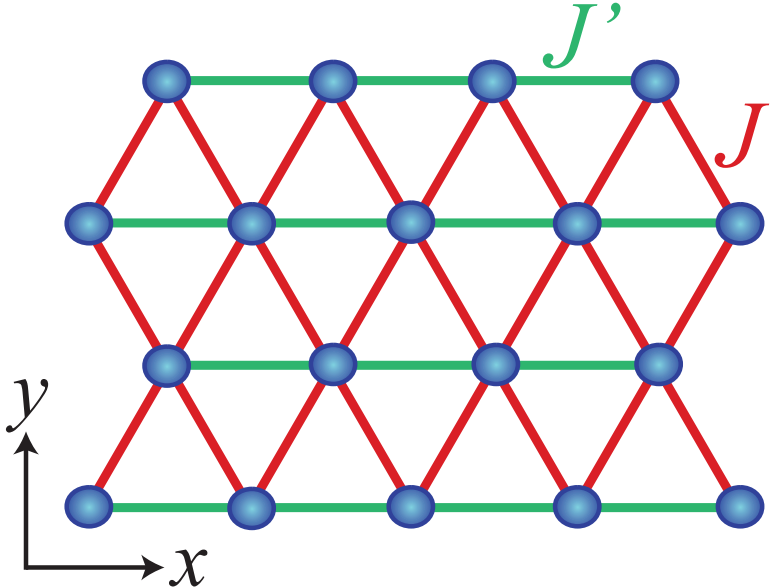}
\caption{Distorted triangular lattice with nearest-neighbor Heisenberg
exchanges $J'$ (on all horizontal bonds) and $J$ (on all other
bonds), representing the geometry of systems examined in this
paper.} \label{lattice}
\end{center}
\end{figure}

As will become clear from our analysis, we can generate distinct
`global' phase diagrams for distinct lattice types and exchange
interactions in two spatial dimensions. We will present a general
method for analyzing these, but will focus on a single lattice
type, found in a number of experimental systems: this is the
distorted triangular lattice shown in Fig.~\ref{lattice}.
Thus we are interested in the $S=1/2$ antiferromagnet, with
SU(2) invariant Heisenberg exchange interactions $J$ and $J'$ illustrated
in Fig.~\ref{lattice}, along with possibly additional longer range
two- or multi-spin exchange interactions which have the same symmetry as Fig.~\ref{lattice}.
A number of limiting cases of this lattice have been examined
earlier, and we will connect
with all of these results:\\
({\em i\/}) for $J' \ll J$ the model becomes essentially
equivalent to the square lattice antiferromagnets considered in
Refs.~\onlinecite{rsl,rsb,senthil}, and our results agree with
these
earlier results in this regime;\\
({\em ii\/}) for $J \ll J'$ we have the quasi-one dimensional
antiferromagnets which have been studied in some detail
by Starykh and Balents \cite{sb1,sb2}; \\
({\em iii\/}) for $J \approx J'$ we have the triangular lattice
antiferromagnets for which our results will connect with those of
Refs.~\onlinecite{rstl,jalabert,sr,sst}; \\
({\em iv\/}) Weihong, McKenzie and Singh \cite{wms} have performed
a series expansion study for the entire range of $J'/J$, and
obtained phases which will also appear in our phase diagrams;\\
({\em v\/}) there have been a number of numerical studies \cite{misguich1,misguich2,mot,sheng} of isotropic
triangular lattice case, $J'=J$, but with an additional four-spin ring exchange interaction, and our theory
will provide candidate phase diagrams for this model; and\\
({\em vi\/}) the non-magnetic phases for $J'=J$ have been modeled by the
quantum dimer model \cite{rk} on the triangular lattice by Moessner and Sondhi\cite{sondhi}, and
our theory will also find their phases.

Experimental examples extend over the full range of parameters for
the lattice in Fig.~\ref{lattice}, and realize a variety of
phases:
\begin{itemize}
\item
A remarkable series of experiments have been carried out by
R.~Kato and collaborators
\cite{kato1,kato2,kato3,kato4,kato5,kato6,kato7} on the organic
Mott insulators X[Pd(dmit)$_2$]$_2$ (for a general review of the
organic compounds, see Ref.~\onlinecite{powell}). Each site of the
lattice in Fig.~\ref{lattice} has a pair of Pd(dmit)$_2$ molecules
carrying charge $-e$ and spin $S=1/2$. X ranges over a variety of
monovalent cations, and the choice of different X allows
experiments over a range of values of $J'/J$. The resulting phase
diagram \cite{kato6} has magnetic order with decreasing critical
temperatures from $T_c  \approx 42$ K to $T_c \approx 15$ K across
the compounds X = Me$_4$P, Me$_4$As, EtMe$_3$As, Et$_2$Me$_2$P,
Et$_2$Me$_2$As, and Me$_4$Sb, as the value of $J'/J$ increases
from $J'/J \approx 0.35$ to $J'/J \approx 0.7$ (there are uncertainties in the
overall scale of $J'/J$, and there are also likely to be significant four-spin
ring exchange terms). The magnetic order is likely of the two-sublattice
N\'eel type \cite{kato6}, although there are no neutron scattering observations
confirming this. The compound with
X = EtMe$_3$Sb has $J'/J \approx 0.85$ has no observable N\'eel
order \cite{kato7} and has been suggested to be near the quantum
critical point \cite{kato6} at which the N\'eel order vanishes.
Finally, the compound \cite{kato4,kato5} with X = EtMe$_3$P has
$J'/J \approx 1.05$ has a ground state with a spin gap and
spontaneous columnar valence bond solid (VBS) order at low $T$.
The VBS order vanishes at a phase transition observed to be at 25
K, and the low $T$ spin gap is measured to be $\approx 40$ K (the
exchange constant $J \approx 250$ K). Thus these series of
compounds appear to realize the N\'eel-VBS transition predicted in
Ref.~\onlinecite{rsl}, and in this paper we will place this
transition in the context of global phase diagrams of models on
the lattice of Fig.~\ref{lattice}. We also note that a N\'eel-VBS
transition as a function of increasing $J'/J$ has also been found
in the series expansion study \cite{wms}.
\item
A separate set of
experiments have been performed on the organic Mott insulators
 $\kappa$-(ET)$_2$Z, which also realize the $S=1/2$ antiferromagnet on the lattice
 in Fig.~\ref{lattice}. The compound with Z = Cu[N(CN)$_2$]Cl has
$J' /J\approx 0.5$, and has a ground state with N\'eel order
\cite{expneel}, as found in the Pd(dmit)$_2$ series above for
small $J'/J$. The organic insulator with Z = Cu$_2$(CN)$_3$ has
$J'/J \approx 1$, and appears to have no antiferromagnetic or VBS
ordering down to the lowest observed temperatures
\cite{kanoda0,kanoda1,kanoda2,yamashita}. This is therefore a
candidate for a spin liquid ground state, whose nature has
been the subject of recent work \cite{mot,palee,qi,z2}. The
bosonic $Z_2$ spin liquid state proposed for this compound in Ref.~\onlinecite{z2}
will appear in our phase diagrams below, and indeed will be
natural point of departure for our entire analysis. We believe the
experimental observation, noted above, of the other phases in our
phase diagram can be regarded as a point of support for our
perspective. We will briefly mention below how $Z_2$ spin liquid
states with fermionic spinons \cite{palee}, and other related
states, can appear in our approach. \item The transition metal
insulator \cite{radu1,radu2} Cs$_2$CuCl$_4$ has $S=1/2$ Cu ions on
the vertices of the triangular lattice in Fig.~\ref{lattice} with
$J \approx J'/3$. The ground state has spiral antiferromagnetic
order, similar to that present in the perfect triangular lattice
($J = J'$), and as will appear in the phase diagrams below. An
approach starting from the quasi one-dimensional limit $J \ll J'$
has been successfully used \cite{sb1,sb2} to describe the spiral
ground state, and also the inelastic neutron scattering spectrum
at high energies.
\end{itemize}

As noted above, our point of departure is a $Z_2$ spin liquid
state. The earliest proposals of such liquids involved BCS-like
states of paired, charge 0, $S=1/2$ particles (`spinons') which
were either bosons \cite{rstl} or fermions \cite{wen1}.
Fluctuations about this state are expressed in terms of a $Z_2$
gauge theory, in which the spinons carry a $Z_2$ electric charge,
and hence the name of the spin liquid. A large number of other
models of $Z_2$ spin liquids have appeared since
\cite{sf,sondhi,kitaev,wen2,freedman,wang,misguich}. We will find
it convenient to begin with $Z_2$ spin liquid in which the
elementary spinons are bosons because it is connected naturally to
a variety of ordered states found experimentally (which we have described
above). We will denote the bosonic spinons by a complex field
$z_\alpha$, where $\alpha = \uparrow, \downarrow$ is a spin index.

Apart from the spinon, the other fundamental elementary excitation
of a $Z_2$ spin liquid is a charge 0 particle carrying $Z_2$
magnetic flux. This particle was pointed out in Ref.~\onlinecite{rstl}, but
its particular importance to the physical properties of $Z_2$ spin
liquids was emphasized by Senthil and Fisher \cite{sf}, who called it a `vison'.
In all cases we shall consider,
it is possible to combine the real visons into complex scalar
fields $v_a$, where $a=1 \ldots N_v$ is an additional flavor index
which depends upon the nature of the underlying lattice. The
visons are bosons, but the spinons and visons have mutual semionic
statistics \cite{rc,sf}. Consequently, by forming bound states of the
bosonic spinons, $z_\alpha$, and the visons, $v_a$, we obtain
$S=1/2$ spinons which are fermions \cite{rc}. This bound state
formation offers a route to extending our analysis to the case of
fermionic spinons, but we shall not comment further on this in the
present paper.

Our starting point is an effective field theory for the spinons,
$z_\alpha$, and the visons, $v_a$, which implements their mutual
semionic statistics. As discussed generally by Freedman {\em et
al.} \cite{freedman}, and implemented more specifically in $Z_2$
spin liquids with fermionic spinons by Kou, Levin, and Wen
\cite{wenlevin}, this mutual statistics can be realized by a
doubled U(1) Chern-Simons (CS) theory. A similar formalism was
also applied to the cuprates, with a mutual statistics between
spin and charge degrees of freedom \cite{kouqi}. To this end, we
introduce 2 U(1) gauge fields, $a_\mu$ and $b_\mu$, and will
consider effective Lagrangians with the following schematic
structure in 2+1 spacetime dimensions
\begin{equation}
\mathcal{L} = \sum_{\alpha = 1}^2\Big\{ |(\partial_\mu - i
a_\mu)z_\alpha|^2  + s_z |z_\alpha|^2 \Bigr\} + \sum_{a=1}^{N_v}
\Bigl\{ |(\partial_\mu - i b_\mu)v_a|^2 + s_v |v_a|^2 \Bigr\} +
\frac{ik}{2\pi} \epsilon_{\mu\nu\lambda}a_\mu\partial_\nu
b_\lambda + \cdots, \label{lcs}
\end{equation}
where $\mu, \nu, \lambda$ = $x,y,\tau$ are spacetime indices, and
$s_z$ and $s_v$ are the primary couplings we will tune to obtain
our global phase diagrams. The integer $k=2$ implements the needed
semionic statistics. The ellipses represent additional terms in
the effective potential for the $z_\alpha$ and $v_a$ which are
constrained by the projective symmetry group (PSG) {\em i.e.\/}
the transformations of the spinons and visons under the symmetries
of the lattice spin Hamiltonian. We will discuss these terms more
carefully when we describe the different PSGs in the body of the
paper.

In passing, we note that supersymmetric versions of the doubled
Chern-Simons theory in Eq.~(\ref{lcs}) have recently been the
focus of intense interest in the string theory literature
\cite{m2a,m2b,m2c,m2d,m2e,m2f}. Their model of interest is
\cite{m2c} a Chern-Simons theory with a U($N$) $\times$ U($N$)
gauge group, with opposite signs for the Chern-Simons term for the
two U($N$)'s, and with matter fields which are bifundamentals in
the U($N$)s. Eq.~(\ref{lcs}) is also precisely of this form with $N=1$: we
can define $c_\mu = a_\mu + b_\mu$ and $d_\mu = a_\mu - b_\mu$,
and then the $c_\mu$ and $d_\mu$ fields have diagonal Chern-Simons
terms with opposite signs, and the $z_\alpha$ and $v_a$ carry
bifundamental charges. The U($N$) $\times$ U($N$) theories have
been argued \cite{m2d,m2e} to be dual to M theory on $AdS_4 \times
S^7/Z_k$, which is reason for the interest. The $N=1$ case with
$\mathcal{N}=4$ supersymmetry has been argued \cite{m2f} to be
exactly dual to a U(1) gauge theory without a Chern-Simons terms (the latter
theory was reviewed in Ref.~\onlinecite{xiss}). The analogs of such
$N=1$ dualities for non-supersymmetric theories are well-known
in the condensed matter literature, and we will discuss
examples in the present paper (see also Ref.~\onlinecite{sf}).

A natural question arises at this point: what are the conditions
under which it is permissible to implement a U(1) CS theory
realization of the $Z_2$ spin liquid, rather than directly in
terms of a $Z_2$ gauge theory? When we are discussing the
topological properties of the ground state, or about single
quasiparticle excitations, there does not appear to be any
obstacle to using a U(1) theory \cite{wenlevin}. However, the
issue becomes more delicate when the excitations proliferate, and
we are considering quantum phase transitions out of the spin
liquid state. This question is discussed further in
Section~\ref{sec:beyond}, where we will find examples of
transitions at which our U(1) CS description fails. However, we
also find cases where it does succeed, and these are the main
focus of this paper. As we will see in Section~\ref{sec:phase},
for these successful cases, because of the constraints of the
lattice PSG, the lowest order terms which break either of the U(1)
gauge invariances of Eq.~(\ref{lcs}) are of eighth order, $\sim
v_a^8$ as in Eq.~(\ref{l8}); their effects are easily
incorporated into our analysis as a soft symmetry breaking. We
mention that the connection between doubled $Z_2$ and U(1) CS
theories has also been discussed by Balents and Fisher in a
different context \cite{bf}.

Crucial to our analysis will be exact results on the low energy
spectrum of $\mathcal{L}$ on a $L \times L$ torus as a function of
$ s_z$ and $s_v$. For the case where both $s_v$ and $s_z$ are
large and positive, both the spinons and visons are gapped, and we
realize a $Z_2$ spin liquid. Here we can integrate out the
$z_\alpha$ and $v_a$, and are left with a pure doubled
Chern-Simons (CS)  gauge theory. This theory was quantized exactly
on a torus in Refs.~\onlinecite{freedman,wenlevin}. The key
variables in this quantization were the fluxes piercing the two
cycles, $C_{x,y}$ of the torus
\begin{equation}
A_i = \oint_{C_i} a_\mu dx_\mu~~,~~B_i = \oint_{C_i} b_\mu d x_\mu .
\label{cycle}
\end{equation}
Given that all the matter fields carry unit $a_\mu$ or $b_\mu$
charges, the $A_i$ and $B_i$ should be regarded as periodic
variables taking values on a circle of circumference $2 \pi$.
After accounting for this periodicity,  The solution of the ground
state of the CS theory was found to be 4-fold degenerate; the
degeneracy appears exponentially fast as $L \rightarrow \infty$,
provided the vison and spinon gaps remain finite. This 4-fold
degeneracy is viewed as an essential characterization of the $Z_2$
spin liquid \cite{rc,rstl}.

The other phases in our phase diagrams appear when we allow one or
both of $s_z$ and $s_v$ to vary to negative values. Then we can
have phase transitions to new phases in which one or both of the
$z_\alpha$ and $v_a$ are ``condensed''. However, the precise
nature of the broken symmetry, if any, is not immediately obvious
in such phases, given the presence of the 2 gauge fields and their
CS term.
\begin{figure}
\begin{center}
\includegraphics[width=5.5in]{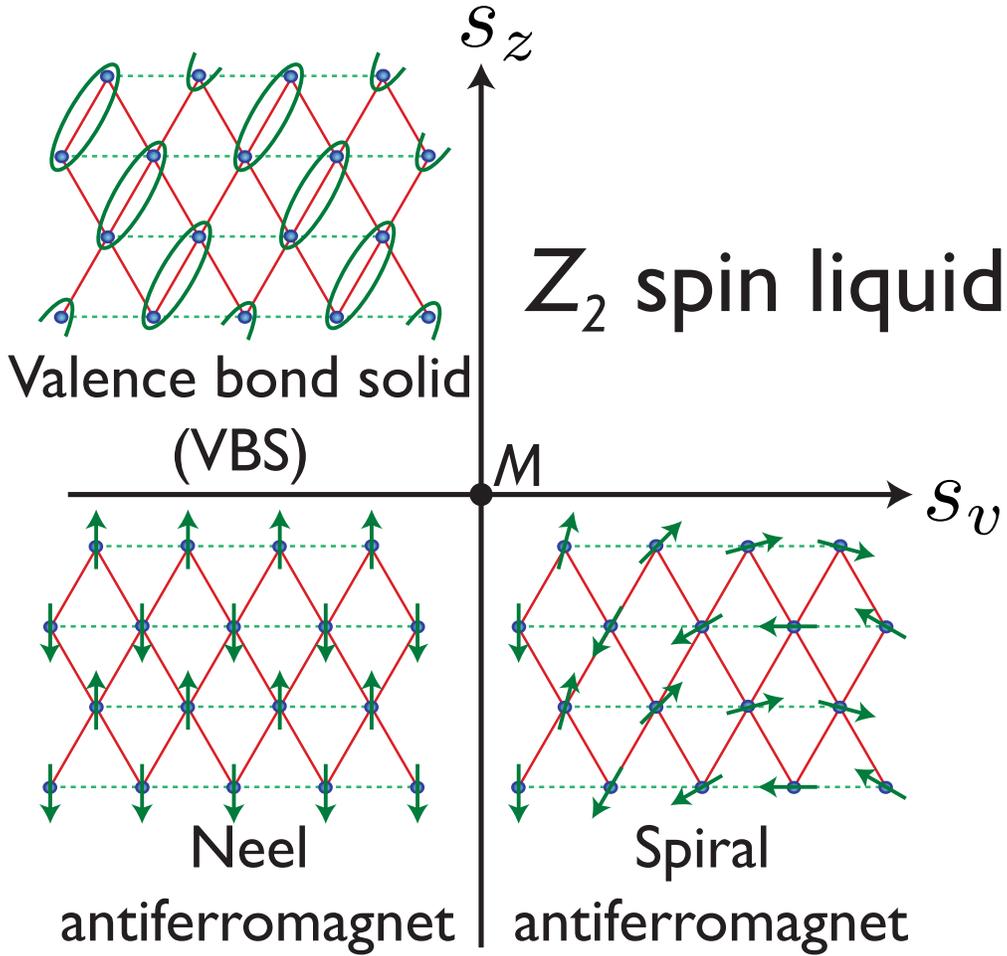}
\caption{Global phase diagram for the case with $N_v=1$ and
a particular model of the spinons (model BIII); a similar phase diagram
appeared in Ref.~\onlinecite{sr}. The N\'eel antiferromagnet is
found in a many materials, and has ordering wavevector ${\bf Q} =
2 \pi (1,0)$. The geometry of the VBS state coincides with that
found\cite{kato4,kato5} in EtMe$_3$P[Pd(dmit)$_2$]$_2$. The spiral
antiferromagnet shown in the figure has an ordering wavevector
${\bf Q} = 2 \pi (1- \epsilon, 0)$ with $\epsilon = 1/6$, small,
as expected for $J'<J$. The experimentally realized spiral state
in Cs$_2$CuCl$_4$ has $J' > J$, and consequently a larger value of
$\epsilon \approx 1/2$. The $Z_2$ spin liquid in this phase
diagram is similar to that proposed in Ref.~\onlinecite{z2} to
explain observations \cite{kanoda0,kanoda1,kanoda2,yamashita} in
$\kappa$-(ET)$_2$Cu$_2$(CN)$_3$. The
transitions have been discussed previously: ({\em i\/}) the
CP$^{1}$ field theory for the N\'eel-VBS transition
\cite{senthil}, ({\em ii\/}) the O(4) field theory for the
spiral-$Z_2$ spin liquid transition \cite{azaria,css,kim}, ({\em
iii\/}) the mean field theory for the spiral-N\'eel transition
\cite{sr}, and ({\em iv\/}) the O(2) field theory for the
VBS-$Z_2$ spin liquid transition \cite{jalabert,sondhi}. All these
field theories are contained in our theory in Eq.~(\ref{lcs}),
which also describes the multicritical point M. } \label{phasesq}
\end{center}
\end{figure}
The purpose of this paper is to describe these new phases and the
associated quantum critical points. Here we note that the order
parameter characterizing these states can be gleaned by carefully
examining the low energy states of $\mathcal{L}$ on a $L \times L$
torus. As an example, consider the state where $s_z$ is large and
negative, and so a saddle point with $z_\alpha \neq 0$ is favored.
By global SU(2) spin symmetry, there are actually an infinite
number of such saddle points along the manifold $|z_\uparrow|^2 +
|z_\downarrow|^2 = $ constant {\em i.e.\/} along $S^3$, the
surface of a sphere in four dimensions. The low energy theory on a
$L \times L$ torus can be expressed as a functional integral over
$S^3$, along with an integral over the gauge fields. We will solve
this quantum theory exactly, and find an ``Anderson tower of
states'' \cite{tower}, with a non-degenerate ground state, and an
infinite sequence of excited levels with energies $\sim 1/L^2$
above the ground state energy. The sequence of excited levels, and
their degeneracies, can be uniquely identified with the quantum
mechanics of a particle moving on $S^3 /Z_2$, with the
co-ordinates of the particle representing the average orientation
of the order parameter across the entire torus. In other words,
the primary effect of the gauge fluctuations is to reduce the
order parameter characterizing the broken symmetry from $S^3$ to
$S^3/Z_2$. It was these same gauge fluctuations which were
responsible for the 4-fold degeneracy in the $Z_2$ spin liquid.
The $S^3/Z_2 \cong SO(3)$ order parameter allows an immediate
identification of the $s_z$ large and negative state: this is the
spiral antiferromagnet, as found in Cs$_2$CuCl$_4$.

A related analysis for the other phases will be found in the body
of the paper, allowing us to construct our phase diagrams. Here we
show in Fig.~\ref{phasesq} the phase diagram found for a case in
which there is only one vison flavor, $N_v = 1$ and for a particular
model of the spinons---we label this theory model BIII.
A mean-field phase
diagram with the same phases appeared already in
Ref.~\onlinecite{sr}, and here we shall show that these phases
follow from the very general considerations outlined above, and
also provide field theories for all the transitions and the
multicritical point $M$. It is encouraging that all the phases
with broken symmetry, which descend from the $Z_2$ spin liquid
with $N_v=1$, correspond precisely to those which have been
experimentally observed so far.

In Section~\ref{sec:cs} we will describe the phase diagram of the
theory in Eq.~(\ref{lcs}) as an abstract field theory, without
reference to any underlying antiferromagnet. The specific spinon
and vison degrees of freedom of the lattice antiferromagnet, and
their possible PSGs, will be identified in Section~\ref{sec:psg}.
The combination of the results of Sections~\ref{sec:cs}
and~\ref{sec:psg} lead to a variety of possible phase diagrams.
These are described in Section~\ref{sec:phase}, and the quantum
phase transitions in Section~\ref{sec:qpt}.
Section~\ref{sec:beyond} will give another semiclassical
perspective on our results, which also identifies the limitations
of the present U(1) CS approach. The concluding
Section~\ref{sec:conc} will make some further remarks on recent
experiments.

\section{Phases of the doubled Chern-Simons theory}
\label{sec:cs}

This section will discuss the phase diagram of the doubled CS
theory, considered here as an abstract field theory. The
interpretation of the phases in terms of the underlying
antiferromagnet requires a more specific knowledge of the PSGs of
the spinons and visons, and these will be considered in subsequent
sections.

For the case where the spinons and visons are gapped, as we have
already noted, we obtain the $Z_2$ spin liquid. The fundamental
property of this theory is the 4-fold degeneracy on a $L \times L$
torus, and this appears as a property of the pure doubled CS
theory \cite{freedman,wenlevin}. We are now interested in moving
into one of the phases where one or both of the spinons and visons
are ``condensed'' and understanding the nature of the broken
symmetry. As in the $Z_2$ spin liquid, we will do this here by
examining the low energy states of the theory on the $L \times L$
torus. We will compute the spectrum of Anderson's \cite{tower}
``tower of states'' with excitation energies $\sim 1/L^2$: the
spectrum of states will allow a unique identification of the
ground state manifold (GSM) associated with the broken global
symmetry.

We will only consider here the case where there is a single spinon
species, $z$, and a single vison species $v$; the generalization
to the multiple species case is straightforward. For the broken
symmetry phases, we need only consider the phases of these complex
fields, and so we will write $z \sim e^{i \theta_z}$ and $v \sim
e^{i \theta_v}$ where the corresponding symmetries are broken.

As a warmup, consider first the case with a single broken U(1)
symmetry, characterized by the U(1) order parameter $e^{i
\theta}$, and no gauge fields. The low energy theory of $\theta$
fluctuations is given by the action
\begin{equation}
\mathcal{S}_\theta = \int d^2 r d \tau \left[ \frac{K_1}{2} (
\partial_\tau \theta)^2 + \frac{K_2}{2} ( \partial_i \theta)^2
\right], \label{e0}
\end{equation}
on a $L \times L$ torus, with $\theta$ and $\theta+2\pi$
identified. The couplings $K_{1,2}$ are two stiffnesses
characterizing the broken symmetry. Because this is a Gaussian
theory, we can make the mode expansion
\begin{equation}
\theta (x, y, \tau) = \theta_0 (\tau) + \frac{2 \pi m x}{L} +
\frac{2 \pi n y}{L} + \frac{1}{L} \sum_{k \neq 0} a_{k} (\tau)
e^{i k \cdot r}
\end{equation}
where $n$ and $m$ are fixed integers (the winding numbers),
$\theta_0$ represents the uniform fluctuation of the order
parameter, and the $a_k$ are the `spin-wave' normal modes.
Inserting this in the action we obtain
\begin{equation}
\mathcal{S}_\theta = 4 \pi^2 (m^2 + n^2) + \int d \tau  \left[
\frac{K_1 L^2}{2} ( \partial_\tau \theta_0 )^2  + \sum_{k \neq 0}
\left( \frac{K_1}{2} (\partial_\tau a_k)^2 + \frac{K_2}{2} k^2
a_k^2 \right) \right]
\end{equation}
From this it is clear that the low-lying states have $m=n=0$. The
$a_k$ harmonic oscillators have energy $\sim k \sim 1/L$, while
the $\theta_0$ mode has energy $\sim 1/L^2$. So for the lowest
states, we put all the $a_k$ oscillators in the ground state, and
we obtain a tower of states with energy
\begin{equation}
E_p = E_0 + \frac{p^2}{2 K_1 L^2}
\label{e1}
\end{equation}
where $p$ is an integer, measuring the angular momentum of the
$\theta_0$ mode.

It is now useful to note that the theory $\mathcal{S}_\theta$ in
Eq.~(\ref{e0}) is exactly dual to U(1) gauge theory with a Maxwell
term, and so the latter should have the same tower of low energy
states. Let us demonstrate this explicitly. First, we decouple the
quadratic terms in Eq.~(\ref{e0}) by an auxiliary current $J_\mu$
\begin{equation}
\mathcal{S}_\theta = \int d^2 r d \tau \left[ \frac{J_\tau^2}{2K_1} +
\frac{J_i^2}{2 K_{2}}  + i J_\mu \partial_\mu \theta
\right], \label{e0a}
\end{equation}
Integrating over $\theta$, we obtain the constraint $\partial_\mu
J_\mu =0$, which we solve by expressing $J_\mu$ in terms of a
`dual' gauge field $a_\mu$:
\begin{equation}
J_\mu = \frac{1}{2 \pi} \epsilon_{\mu\nu\lambda} \partial_\nu
a_\lambda.  \label{e0b}
\end{equation}
The normalization of $1/(2 \pi)$ is chosen so that periodicity of
the flux variables $A_i$ in Eq.~(\ref{cycle}) with period $2 \pi$
is equivalent to the periodicity in the angular variable $\theta
\rightarrow \theta + 2 \pi$. Now inserting Eq.~(\ref{e0b}) into
(\ref{e0a}) we obtain the U(1) gauge theory dual to
$\mathcal{S}_\theta$:
\begin{equation}
\mathcal{S}_a = \int d^2 r d \tau \left[ \frac{1}{8 \pi^2 K_2} (
\partial_\tau a_i)^2 + \frac{1}{ 8 \pi^2 K_1} ( \partial_x a_y -
\partial_y a_x)^2 \right] \label{e0c}
\end{equation}
where we have chosen the temporal gauge with $a_\tau = 0$.  An
important consequence of the periodicity in $A_i$ variables is
that the flux piercing the torus, $\int d^2r (\partial_x a_y -
\partial_y a_x)$ must be an integer multiple of $2 \pi$. We can
see this by moving the contour $C_i$ in Eq.~(\ref{cycle}) across
the entire length of the torus: the change in the line integral
upon returning to the initial position must be an integer multiple
of $2 \pi$, and this change is equal by Stokes theorem to the flux
piercing the torus. Thus the Hilbert space of $\mathcal{S}_a$ breaks apart
into distinct sectors with total flux $2 \pi p$, where $p$ is an
integer. Within each sector, the ground state has zero photons
(which are the dual of the spin waves), and has $\langle
\partial_x a_y - \partial_y a_x \rangle= (2 \pi p)/L^2$. So the
lowest energy state in each sector is
\begin{equation}
E_p = E_0 + \frac{L^2}{8 \pi^2 K_1} \left( \frac{2 \pi p}{L^2}
\right)^2 = E_0 + \frac{p^2}{2 K_1 L^2} \label{e5}
\end{equation}
which is the same as the spectrum of $S_\theta$ inEq.~(\ref{e1}).
This verifies the equivalence of Eq.~(\ref{e0}) and (\ref{e0c}).

With these preliminaries out of the way, let us return to our
doubled CS theory with one spinon and one vison. Consider the
phase where the spinon is condensed and the vison is gapped, so
$s_v \gg 0$ and $s_z \ll 0$. Here we can simply integrate out the
vison, and are left with the following low energy theory for
$\theta_z$ and the U(1) gauge fields
\begin{equation}
\mathcal{S}_z = \int d^2 r d \tau \left[ \frac{K_1}{2} (
\partial_\tau \theta_z - a_\tau )^2 + \frac{K_2}{2} ( \partial_i
\theta_z + a_i )^2 + \frac{i k}{2 \pi} a_\mu
\epsilon_{\mu\nu\lambda} \partial_\nu b_\lambda \right] \label{e3}
\end{equation}
We can always choose the gauge $\theta_z =0$ (and $A_\tau=0$). In
this gauge, the integral over $a_\mu$ is an ordinary Gaussian.
Performing this integral, we obtain the action
\begin{equation}
\mathcal{S}_z = \int d^2 r d \tau \left[ \frac{k^2}{8 \pi^2 K_2} (
\partial_\tau b_i)^2 + \frac{k^2}{ 8 \pi^2 K_1} ( \partial_x b_y -
\partial_y b_x)^2 \right] \label{szcs}
\end{equation}
Comparing this with the spectrum of $\mathcal{S}_a$ in
Eq.~(\ref{e5}), we obtain the low-lying states
\begin{equation}
E_p = E_0 + \frac{k^2 p^2}{2 K_1 L^2}
\end{equation}
This shows that the theory $\mathcal{S}_z$ in Eq.~(\ref{e3}) is
equivalent to the $U(1)$ scalar theory in Eq.~(\ref{e0}) but with
the periodicity $\theta \equiv \theta + 2 \pi/k$. In other words,
the GSM of this phase has been modified by the gauge fluctuations
from $S^1$ to $S^1/Z_k$. Alternatively stated, the broken symmetry
of the ground state is associated with distinct values of the
composite field $z^k$.

It is useful to have another perspective on the above result by an
alternative analysis of the theory $\mathcal{S}_z$ in
Eq.~(\ref{e3}). This analysis begins by `undualizing' the gauge
field $b_\mu$ into a dual scalar $\theta_b$. For this we
introduce, as in Eq.~(\ref{e0b}), the current $J^b_\mu =
\epsilon_{\mu\nu\lambda} \partial_\nu b_{\lambda} /(2 \pi)$ and
impose the constraint $\partial_\mu J^b_\mu = 0$ by a Lagrange
multiplier $\theta_b$; this modifies $\mathcal{S}_z$ in
Eq.~(\ref{e3}) to
\begin{equation}
\mathcal{S}_z = \int d^2 r d \tau \left[ \frac{K_1}{2} (
\partial_\tau \theta_z - a_\tau )^2 + \frac{K_2}{2} ( \partial_i
\theta_z + a_i )^2 + i J_\mu^b (k a_\mu - \partial_\mu \theta_b) +
\frac{J_\mu^{b2}}{2 \widetilde{K}} \right] \label{e3a}
\end{equation}
The last term is a useful regularization, and the original theory
in Eq.~(\ref{e3}) is obtained in the limit $\widetilde{K}
\rightarrow \infty$. Now we perform the integral over $J_\mu^b$
and obtain the theory
\begin{equation}
\mathcal{S}_z = \int d^2 r d \tau \left[ \frac{K_1}{2} (
\partial_\tau \theta_z - a_\tau )^2 + \frac{K_2}{2} ( \partial_i
\theta_z + a_i )^2 + \frac{\widetilde{K}}{2} (k a_\mu -
\partial_\mu \theta_b)^2 \right] \label{e3b}
\end{equation}
In the limit $\widetilde{K} \rightarrow \infty$, we see that we
must have $a_\mu = (1/k) \partial_\mu \theta_b$. However,
$\theta_b$ is a variable periodic under $\theta_b \rightarrow
\theta_b + 2 \pi$, and hence the periodic flux variables $A_i$ in
Eq.~(\ref{cycle}) can only take the values
\begin{equation}
A_i = \frac{2 \pi p_i}{k}
\end{equation}
where the $p_i$ are integers. In other words, U(1) gauge field
$a_\mu$ has been reduced to a $Z_k$ gauge field. Thus $e^{i k
\theta} \sim z^k$ is gauge invariant, and this explains our
results above on the distinct values of $z^k$ identifying distinct
ground states.

We have now completed our discussion of the state where the spinon
is condensed and the vison is gapped ($s_v \gg 0$ and $s_z \ll
0$). Clearly, the complementary phase where the vison is condensed
and the spinon is gapped ($s_v \ll 0$ and $s_z \gg 0$) is amenable
to a parallel treatment, with complementary results and a $v^k$
order parameter.

Finally, let us consider the case where both the visons and
spinons are condensed, $s_v \ll 0$ and $s_z \ll 0$. In this case,
we can see in the gauge $\theta_z=0$ and $\theta_v=0$ that both
fields $a_\mu$ and $b_\mu$ are fully gapped. So there is a unique
ground state, and no other excited states whose energy vanishes as
$L \rightarrow \infty$.

\begin{figure}
\begin{center}
\includegraphics[width=4.5in]{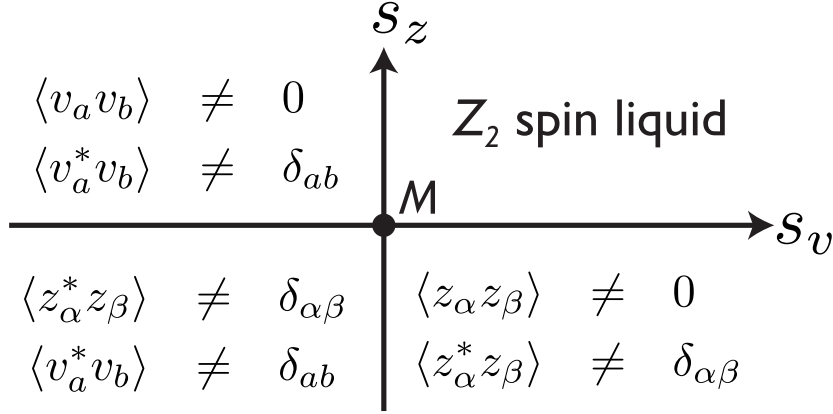}
\caption{Schematic phase diagram of the doubled CS theory in
Eq.~(\ref{lcs}) for $k=2$. All non-zero order parameters which
characterize the broken symmetry in each phase are shown.}
\label{phasecs}
\end{center}
\end{figure}
We can now generalize these results to the cases with multiple
flavors of visons and spinons, and the results are summarized in
Fig.~\ref{phasecs} for $k=2$. The flavor indices simply tag along
for the order parameters involving a $k$-fold composites of
spinons or visons, which are invariant under the $Z_k$ gauge
transformations. However, there are now also additional
gauge-neutral order parameters possible, like $z^\ast_\alpha
z_\beta$, which were absent for the single flavor case, because
they were not associated with any broken symmetry.

Armed with the results in Fig.~\ref{phasecs}, and with a knowledge
of the microscopic PSGs of the spinons and visons (which are
described next in Section~\ref{sec:psg}), we can easily deduce the
physical characteristics of the phases of a variety of
antiferromagnets.

\section{Spinons and visons}
\label{sec:psg}

A rich variety of spinon and vison operators can be defined for
$S=1/2$ antiferromagnets on the lattice in Fig.~\ref{lattice}, and
we shall not attempt any complete classification. Clearly, the
choice depends sensitively on the details of the microscopic
Hamiltonian. However, in previous semiclassical analyses, a few
natural choices have emerged for different limiting values of
$J'/J$. We will describe these below, and show how they can fit
together in a doubled CS theory like Eq.~(\ref{lcs}).

The essential characteristics of the spinons and visons will be
their transformations under the symmetry operations of the
underlying spin model. These symmetries are the lattice
translations $T_1$ and $T_2$, the lattice reflections $P_x$ and
$P_y$, and time reversal, $T$:
\begin{eqnarray}
T_1 &:& (x,y) \rightarrow (x+1,y) \nonumber \\
T_2 &:& (x,y) \rightarrow (x+1/2,y+\sqrt{3}/2) \nonumber \\
P_x &:& (x,y) \rightarrow (-x,y) \nonumber \\
P_y &:& (x,y) \rightarrow (x,-y) \nonumber \\
T &:& t \rightarrow -t
\end{eqnarray}
Spin rotation is also a symmetry, and is easily
implemented by contracting the spinor indices.

The following subsections will consider the PSGs of spinons and
visons in turn.

\subsection{Spinons}
\label{sec:spinons}

One natural model of spinons appears upon describing the spiral
ground state of the triangular lattice antiferromagnet in terms of
Schwinger bosons. So we write the spin operator on the lattice
sites as $\vec{S} = b^\dagger \vec{\sigma} b/2$, where
$\vec{\sigma}$ are the Pauli matrices. For the Schwinger bosons we
make the following low energy expansion \cite{ssrmp} in terms of the spinon
fields, $z_\alpha$: \beqn b_\alpha \sim z_\alpha \exp(i {\bf
Q}\cdot {\bf r}/2) + i \epsilon_{\alpha\beta}z^\ast_\beta
\exp(-i{\bf Q}\cdot{\bf r}/2), \label{spinon} \eeqn From this
parameterization we can then deduce the following expression for
the spin operators \beqn \vec{S} &=& \vec{n}_1 \cos({\bf Q}\cdot
{\bf r}) + \vec{n}_2 \sin({\bf Q}\cdot {\bf r}), \cr \vec{n}_1 &=&
\mathrm{Re}[z^t \sigma^y \vec{\sigma} z], \ \ \vec{n}_2 =
\mathrm{Im}[z^t \sigma^y \vec{\sigma} z], \cr \vec{n}_3 &=&
\vec{n}_1 \times \vec{n}_2 = z^\dagger \vec{\sigma} z. \label{szq}
\eeqn We observe that ${\bf Q}$ is the ordering wavevector of the
spiral. So for $J' \approx J$, we expect ${\bf Q} \approx (2
\pi/3, 0)$. For Cs$_2$CuCl$_4$, which has $J' \approx 3 J$, we
have ${\bf Q} \approx (\pi, 0)$. Finally, in the square lattice
limit, $J' \ll J$, we have ${\bf Q} \approx (2 \pi, 0)$. Thus we
expect ${\bf Q}$ to increase monotonically from $(\pi, 0)$ to $(2
\pi, 0)$ with decreasing $J'/J$. Also note that $\vec{n}_{1,2,3}$
are three mutually orthogonal vectors.

The parameterization in Eq.~(\ref{szq}) allows us to deduce the
PSG of the $z_\alpha$. These $z_\alpha$ spinons will couple
minimally to the $a_\mu$ gauge field, and so there is a natural
implied PSG for the $a_\mu$. We call the resulting PSG of spinons
as model A:
\begin{eqnarray}
&& \mbox{\underline{Spinons, Model A}} \\
T_{1} &:& z \rightarrow e^{iQ_x /2}z , \ \  a_\mu \rightarrow
a_\mu \cr T_{2} &:& z \rightarrow e^{iQ_x / 4} z , \  \ a_\mu
\rightarrow a_\mu \cr P_x &:& z_\alpha \rightarrow
\epsilon_{\alpha\beta}z^\ast_\beta, \  \ a_x \rightarrow a_x, \  \
a_y \rightarrow -a_y, \  \ a_t \rightarrow -a_t \cr P_y &:& z
\rightarrow z,  \  \ a_x \rightarrow a_x, \  \ a_y \rightarrow
-a_y, \  \ a_t \rightarrow a_t \cr T &:& z \rightarrow i z^\ast, \
\ a_\mu \rightarrow a_\mu.
\label{psga}
\end{eqnarray}
We note that under this model A PSG, $\vec{n}_{1,2}$ are odd under
time reversal, while $\vec{n}_3$ is even. From this, and the
representation in Eq.~(\ref{szq}), we deduce that $\vec{n}_3$ is a
spin nematic order parameter.

The Model A spinons are natural for $J' \gtrsim J$, where the
spiral order is likely to be present. However as we approach the
square lattice limit with $J' \rightarrow 0$, there is a finite
range of small $J' < J$ over which we expect that ${\bf Q}$ is
pinned exactly at $(2 \pi, 0)$, and we have the conventional 2
sublattice N\'eel order appropriate for the square lattice. In
this case $\sin ({\bf Q} \cdot {\bf r}) =0$ identically, and
$\cos({\bf Q} \cdot {\bf r}) = \cos (2 \pi x) = (-1)^{2x}$. In
this limit, we can define another model of spinons which appeared
in previous theories of square lattice antiferromagnets
\cite{rsl,rsb}. We map $z_\alpha \rightarrow (z_\alpha + i
\epsilon_{\alpha\beta} z_\beta^\ast )/\sqrt{2}$ and then find that
Eq.~(\ref{szq}) is replaced by \beqn \vec{S} = \vec{m}_1 (-1)^{2x}
~,~\vec{m}_1 = z^\dagger \vec{\sigma}  z,  \label{szqb} \eeqn A
significant difference from Eq.~(\ref{szq}) is that now the U(1)
gauge invariance associated with $a_\mu$ is explicit, because the
representation (\ref{szqb}) is invariant under the gauge
transformation $z_\alpha \rightarrow z_\alpha e^{i \theta}$. We
label these spinons model B, and they also have mappings under the
square lattice space group, which we can deduce from
Eq.~(\ref{szqb}) to be:
\begin{eqnarray}
&& \mbox{\underline{Spinons, Model B}} \\
T_{1} &:& z \rightarrow -z  , \ \  a_\mu \rightarrow a_\mu \cr
T_{2} &:& z_\alpha \rightarrow - \epsilon_{\alpha\beta}
z_\beta^\ast , \  \ a_\mu \rightarrow -a_\mu \cr P_x &:& z
\rightarrow i z , \  \ a_x \rightarrow -a_x, \  \ a_y \rightarrow
a_y, \  \ a_t \rightarrow a_t \cr P_y &:& z \rightarrow z,  \  \
a_x \rightarrow a_x, \  \ a_y \rightarrow -a_y, \  \ a_t
\rightarrow a_t \cr T &:& z_\alpha \rightarrow
\epsilon_{\alpha\beta}z_\beta, \ \ a_\mu \rightarrow -a_\mu.
\label{psgmodelb}
\end{eqnarray}
It is now clear that under the model B PSG, $\vec{m}_1$ is odd
under time-reversal. In the state where the spinons are condensed
and the visons are gapped, we see from Fig.~\ref{phasecs} that we
also have two additional vectors in spin space which characterize
the broken symmetry in the ground state (analogous to the 3
vectors found in model A):
\begin{equation}
\vec{m}_2 + i \vec{m}_3 = z^t
\sigma^y \vec{\sigma} z
\label{m2m3z}
\end{equation}
These vectors do not appear in the present expression for the spin
operator in Eq.~(\ref{szqb}), and so their physical interpretation
is not yet clear. Let us, therefore, compute the PSG of these
vectors:
\begin{eqnarray}
&& \mbox{\underline{Model B}} \\
T_{1} &:& \vec{m}_{2,3} \rightarrow \vec{m}_{2,3} \cr T_{2} &:&
\vec{m}_2 \rightarrow \vec{m}_2 \ \ \vec{m}_3 \rightarrow -
\vec{m}_3 \cr P_x &:&  \vec{m}_{2,3} \rightarrow -\vec{m}_{2,3}
\cr P_y &:& \vec{m}_{2,3} \rightarrow \vec{m}_{2,3} \cr T &:&
\vec{m}_2 \rightarrow \vec{m}_2 \ \ \vec{m}_3 \rightarrow -
\vec{m}_3
\end{eqnarray}
From these properties it is evident that we can identify
$\vec{m}_2$ as the central axis (in spin space) about which the
spins are precessing in the spiral antiferromagnet shown in
Fig.~\ref{phasesq}. We will present a more detailed analysis in
Section~\ref{phasediaB} which shows how the spiral state emerges for
model B spinons.
This
identification is also consistent with the analysis of this phase
in Ref.~\onlinecite{sr}, where the spiral phase
was induced by the condensation of a charge 2 Higgs scalar---the
vison is dual to this scalar, and the gapping of the vison is
equivalent to the condensation of the Higgs scalar.

\subsection{Visons}
\label{sec:visons}

\begin{figure}
\begin{center}
\includegraphics[width=4in]{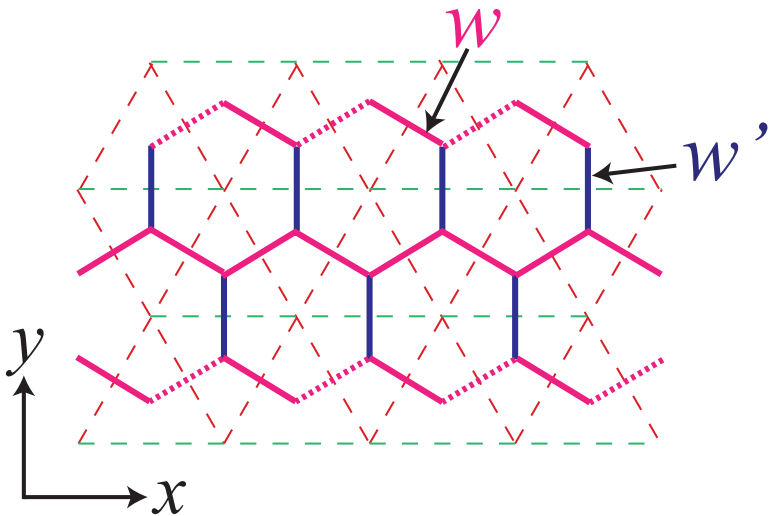}
\caption{The dual Ising model describing the vison dynamics in the
dual honeycomb lattice. All the vertical bonds have hopping
amplitude $w^\prime$, all the other bonds have hopping amplitude
$w$. Note that a small $J'/J$ implies a large $w'/w$, and vice versa.
Besides the lattice anisotropy, there is a $\pi-$flux through
every hexagon, which frustrates the vison kinetics; this flux is implemented
by changing the sign of the dotted $w$ bonds.}
\label{dlattice}
\end{center}
\end{figure}

In the simplest models of visons \cite{jalabert,sf,sondhi}, we
take real particles hopping on the sites of the dual lattice,
subject to a flux of $\pi$ around every site of the direct
lattice. In other words, the vison is the Ising field of an Ising
model on the dual lattice, with exchange couplings chosen so that
every plaquette surrounding a direct lattice site is frustrated.
For the antiferromagnet on the lattice in Fig.~\ref{lattice}, the
Ising model resides on the dual lattice shown in Fig.
\ref{dlattice}. Because of the dual relation between the couplings
of the antiferromagnet, and the model of the visons, we expect
that $w'/w$ decreases as $J'/J$ increases.

It is a relatively straightforward matter to obtain the spectrum
of such a particle moving on the lattice in Fig.~\ref{dlattice}.
Because the kinetics of the vison is frustrated by the background
spinon charge on every site, the product of the hopping amplitudes
on the six links around each hexagon is -1. We are free to choose
one of the links to be negative, and our convention is shown in
Fig.~\ref{dlattice}. The spectrum has multiple minima in the
Brillouin zone, and we introduce a real vison field for each such
minimum in the spectrum. This procedure parallels that carried out
in obtaining the multiple vortex flavors in
Refs.~\onlinecite{courtney} and \onlinecite{bbbss}, but with
modification that we are considering real vison fields and not
complex vortex fields.

For the general set of parameters for the lattice in
Fig.~\ref{dlattice}, we find that there are either 4 or 2 minima
of the vison dispersion lattice in the Brillouin zone. The 4
minima occur near $J' \approx J$, and so are appropriate for the
triangular lattice limit; indeed for $J'=J$ these minima co-incide
with those found by Moessner and Sondhi \cite{sondhi}. For $J'/J$
small ($w'/w$ large), near the square lattice limit, we find only
2 minima.

Let us begin by considering the 4 minima case. These are at
momenta of the form \beqn {\bf q}_1 &=& (q_{1x}, \pi/2 ), \cr {\bf
q}_2 &=& (\pi - q_{1x}, \pi/2 ), \cr {\bf q}_3 &=& ( - q_{1x}, -
\pi/2 ) , \cr {\bf q}_4 &=& (q_{1x} - \pi, - \pi/2 ). \eeqn There
are two choices to combine these four real minima to two complex
minima, which will then correspond to the complex vison fields
$v_a$, with $a=1,2$. The first choice, which we call model I, is:
\beqn v_1: \ {\bf q}_1 &=& (q_{1x}, \pi/2 ), \cr v_2: \ {\bf q}_2
&=& (\pi - q_{1x}, \pi/2 ), \cr v_1^\ast : \ {\bf q}_3 &=& ( -
q_{1x}, - \pi/2 ) , \cr v_2^\ast : {\bf q}_4 &=& (q_{1x} - \pi, -
\pi/2 ),  \eeqn while model II is: \beqn v_1: \ {\bf q}_1 &=&
(q_{1x}, \pi/2 ), \cr v^\ast_2: \ {\bf q}_2 &=& (\pi - q_{1x},
\pi/2 ), \cr v_1^\ast : \ {\bf q}_3 &=& ( - q_{1x}, - \pi/2 ) ,
\cr v_2 : {\bf q}_4 &=& (q_{1x} - \pi, - \pi/2 ). \eeqn

These two choices lead to 2 models for the vison PSGs: \beqn &&
\mbox{\underline{Visons, Model I}} \cr T_{1} &:&  v_1 \rightarrow
e^{iq_{1x}}v_1 , \ \ v_2 \rightarrow e^{i\pi - i q_{1x}}v_2, \ \
b_\mu \rightarrow b_\mu \cr T_{2} &:&  v_1 \rightarrow e^{i\theta}
v_2^\ast , \ \  v_2 \rightarrow -i e^{-i\theta}v_1^\ast, \ \ b_\mu
\rightarrow - b_\mu \cr P_x &:&  v_1 \rightarrow e^{- i\gamma}
v_1^\ast, \ \ v_2 \rightarrow - e^{i\gamma} v_2^\ast , \ \ b_x
\rightarrow b_x, \ \ b_y \rightarrow -b_y, \ \ b_t \rightarrow
-b_t \cr P_y &:&  v_1 \rightarrow v_2^\ast,  \ \ v_2 \rightarrow
v_1^\ast, \ \ b_x \rightarrow -b_x, \ \ b_y \rightarrow b_y, \ \
b_t \rightarrow -b_t \cr T &:&  v_a \rightarrow v_a^\ast, \ \
b_\mu \rightarrow b_\mu \eeqn where $\gamma$ and $\theta$ are two
incommensurate angles.

The second model for the vison PSG is \beqn &&
\mbox{\underline{Visons, Model II}} \cr T_{1} &:&  v_1 \rightarrow
e^{iq_{1x}}v_1 , \ \ v_2 \rightarrow e^{- i\pi + i q_{1x}}v_2,  \
\ b_\mu \rightarrow b_\mu, \cr T_{2} &:&  v_1 \rightarrow
e^{i\theta} v_2 , \ \ v_2 \rightarrow  i e^{i\theta}v_1, \ \ b_\mu
\rightarrow b_\mu, \cr P_x &:&  v_1 \rightarrow e^{- i\gamma}
v_1^\ast, \ \ v_2 \rightarrow - e^{- i\gamma} v_2^\ast , \ \ b_x
\rightarrow b_x, \ \ b_y \rightarrow -b_y, \ \ b_t \rightarrow
-b_t \cr P_y &:& v_1 \rightarrow v_2,  \ \ v_2 \rightarrow v_1, \
\ b_x \rightarrow b_x, \ \ b_y \rightarrow -b_y, \ \ b_t
\rightarrow b_t\cr T &:& v_a \rightarrow v_a^\ast, \ \ b_\mu
\rightarrow b_\mu. \eeqn where again $\gamma$ and $\theta$ are two
incommensurate angles when the lattice is distorted. For an
isotropic triangular lattice, $\theta = -\pi/12$, $\gamma =
\pi/6$, and they will continuously evolve to vison model III by
tuning the distortion of the lattice.

Finally, let us move to the case where the four minima in the
vison band merge to two. In model I, $v_1$ and $v_2$ merge
together, while in model II $v_1$ and $v_2^\ast$ merge together.
But $v_1$ and $v_2^\ast$ carry opposite gauge charges in the CS
theory, and so this merger violates the gauge invariance. So if we
want to evolve smoothly from four minima to two minima, we have to
take model I.

After the merger, the two minima are located at $(\pi/2, \pi/2)$
and $(-\pi/2, -\pi/2)$. So we have just to use one component
complex vison $v$, and its PSG leads to \beqn &&
\mbox{\underline{Visons, Model III}} \cr T_{1} &:&  v \rightarrow
iv , \ \ b_\mu \rightarrow b_\mu \cr T_{2} &:&  v \rightarrow
-e^{3 \pi i /4} v^\ast , \ \ b_\mu \rightarrow - b_\mu \cr P_x &:&
v \rightarrow -iv^\ast , \ \ b_x \rightarrow b_x, \ \ b_y
\rightarrow -b_y, \ \ b_t \rightarrow -b_t \cr P_y &:&  v
\rightarrow v^\ast , \ \ b_x \rightarrow -b_x, \ \ b_y \rightarrow
b_y, \ \ b_t \rightarrow -b_t \cr T &:&  v \rightarrow v^\ast, \ \
b_\mu \rightarrow - b_\mu \eeqn

Finally, we consider the nature of
the vison order parameters $v_a v_b$ and $v_a^\ast v_b$ which
appear in the phases in Fig.~\ref{phasecs}.

The simplest case is model III, with only one complex vison $v$,
in which case the only non-trivial order parameter is $v^2$. From
the PSG, we see that $v^2$ is the square lattice VBS order
parameter, associated with the VBS state shown in
Fig.~\ref{phasesq}. Using the definitions
$V_{\overline{x},\overline{y}}$ for this order parameter in
Ref.~\onlinecite{senthil}, we have $V_{\overline{x}} \sim
\cos(\phi + \pi/4)$, $V_{\overline{y}} \sim \cos(\phi - \pi/4)$
where $v^2 = \exp(i\phi)$. Here (and henceforth), the axes
$\overline{x}$, $\overline{y}$ refer to the principle axes of the
``square'' lattice formed by the $J$ bonds in Fig.~\ref{lattice}.

The vison order parameters for the other models also describe VBS
orders but of a different nature. The vison operator $v_a$ is
subject to a $Z_2$ gauge invariance, therefore the physical VBS
order parameter should always be bilinear of $v_a$. There are in
total 15 independent bilinear of $v_a$, and the detailed VBS
pattern drive by vison proliferation depends on the Hamiltonian of
visons, which will be discussed in the next section.

\section{Phase diagrams}
\label{sec:phase}

Now we turn to the crucial question of combining the spinon and
vison PSGs in Section~\ref{sec:psg} into consistent theories of
the form in Eq.~(\ref{lcs}). We will denote the resulting theories
by an obvious notation {\em i.e.} the theory BIII has spinons
under model B and visons under model III.

In principle, there are now 6 possible theories, AI, AII, AIII,
BI, BII, and BIII, and associated phase diagrams. To establish the
consistency of these theories, we have to examine the
transformation of the CS term under the respective spinon and
vison PSGs. The results of such an analysis are summarized in
Table~\ref{tablecs} for all theories.
\begin{table}
\begin{spacing}{1.5}
\centering
\begin{tabular}{||c||c|c|c|c||} \hline\hline
~~~~~~ &   BI and BIII & AI and AIII & ~~~~AII~~~~ & ~~~~BII~~~~ \\
 \hline\hline
$T_1$ & $+$ & $+$ & $+$ & $+$
\\ \hline
$T_2$ & $+$ & $-$ & $+$ & $-$ \\ \hline
$P_x$ & $+$ & $-$ & $-$ & $+$ \\ \hline
$P_y$ & $+$ & $+$ & $-$ & $-$
\\ \hline
$T$ & $+$ & $-$ & $-$ & $+$
\\ \hline\hline
\end{tabular}
\end{spacing}
\caption{Transformation of the mutual Chern Simons term under the
space group operations for the various theories. The CS term
changes its overall sign, as indicated.} \label{tablecs}
\end{table}
We see that under theories BI and BIII the Chern Simons term is
strictly invariant, and so these theories are clearly consistent.
For the remaining theories, the
overall form of the CS term remains invariant,
but some of the transformations do lead to a
change in sign of the CS term. However, the role of the CS term
here for $k=2$ is only to implement a mutual semionic phase of
$\pi$, and this is invariant under the sign change. Equivalently, we
are free to define the vison at momentum $\mathbf{Q}$ to be either $v$
or $v^\ast$, which means that in the system there should be
particle-hole symmetry of vison $i.e.$ on average the spinons see
zero flux. This particle-hole symmetry corresponds to the free
choice of the sign of gauge charge of vison, and leads to the
freedom of the sign of the mutual CS term.

\subsection{Model AI}

\begin{figure}
\begin{center}
\includegraphics[width=3.5in]{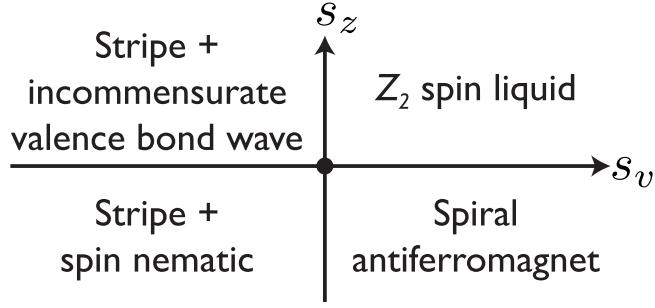}
\caption{Phase diagram of model AI with $g>0$. The stripe order is
illustrated in Fig.~\ref{figstripe}, while the spiral
antiferromagnet is as in Fig.~\ref{phasesq}.} \label{phaseAIp}
\end{center}
\end{figure}

\begin{figure}
\begin{center}
\includegraphics[width=3.5in]{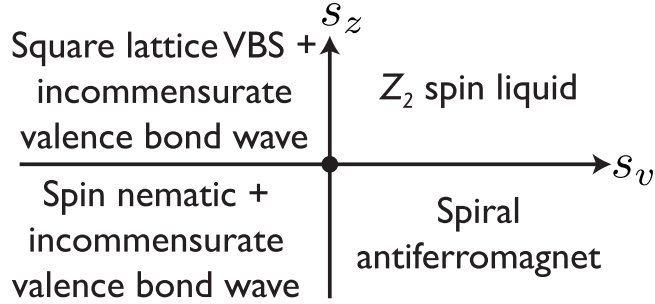}
\caption{Phase diagram of model AI with $g<0$. The square lattice
VBS state is as illustrated in Fig.~\ref{phasesq}.}
\label{phaseAIm}
\end{center}
\end{figure}

The Lagrangian should be invariant under all the symmetry and PSG
transformations, which in general takes the form: \beqn
\mathcal{L} &=& \sum_{\alpha = 1}^2\Big\{ |(\partial_\mu - i
a_\mu)z_\alpha|^2  + s_z |z_\alpha|^2 \Bigr\} + \sum_{a=1}^{2}
\Bigl\{ |(\partial_\mu - i b_\mu)v_a|^2 + s_v |v_a|^2 \Bigr\} +
\frac{ik}{2\pi} \epsilon_{\mu\nu\lambda}a_\mu\partial_\nu
b_\lambda \cr\cr &+& u_z
\left(\sum_{\alpha=1}^2|z_\alpha|\right)^2 + u_v
\left(\sum_{a=1}^2|v_a|\right)^2 + g|v_1|^2|v_2|^2 + \cdots
\label{CSAI} \eeqn Let us first identify all the symmetries of
this Lagrangian. The U(1) gauge symmetries associated with gauge
field $a_\mu$ and $b_\mu$, correspond to two global U(1)
symmetries U(1)$_a\times$ U(1)$_b$ in the dual picture, which lead
to the conservation of gauge fluxes. Through the mutual CS term,
the gauge flux of $a_\mu$ is attached with the vison number, and
the gauge flux of $b_\mu$ is attached with the spinon number. On
top of the U(1) gauge symmetries, the global symmetry of this
Lagrangian to the fourth order of $z_\alpha$ and $v_a$ is
$\mathrm{SU(2)_{spin} \times U(1) \times} Z_2$. The U(1) symmetry
corresponds to the U(1) transformation on vison bilinear $v_1^\ast
v_2$; the $Z_2$ symmetry corresponds to interchanging $v_1$ and
$v_2$, which physically can be understood as the reflection
symmetry $\mathrm{P_y}$. If the lattice is an undistorted
triangular lattice, $g = 0$, and the vison doublet enjoys an
enlarged SU(2) flavor symmetry in this mutual CS theory (or an
O(4) symmetry in a theory with only vison). The ellipses in Eq.
(\ref{CSAI}) include terms no less than sixth order of $v_a$,
which may introduce higher order anisotropy. In the distorted
triangular lattice, the lowest order term which breaks the
symmetries in Eq. (\ref{CSAI}) is at the eighth order: \beqn
\mathcal{L}_8 = g_8 (v_1v_2)^4 + \mbox{H.c.} \label{l8}\eeqn In
the undistorted lattice, the lowest order symmetry breaking term
is at the sixth order. If only the terms below fourth order are
considered, we can minimize the Lagrangian in Eq.~(\ref{CSAI})
with tuning parameter $s_z$ and $s_v$, and obtain the phase
diagrams in Fig.~\ref{phaseAIp} and~\ref{phaseAIm}, with the
phases described in the following subsections.

\subsubsection{The $Z_2$ spin liquid and spiral phases}

The phase with both the spinons and visons gapped out ($s_z > 0$,
$s_v > 0$) is the $Z_2$ spin liquid, as was discussed in
Sections~\ref{sec:intro} and~\ref{sec:cs}, has four-fold
topological degeneracy on a compact torus. The phase with visons
gapped and spinons condensed ($s_v > 0$, $s_z < 0$), is the
incommensurate spin spiral state, with wave vector $\mathbf{Q}$.
With the vison gapped, the gauge field $b_\mu$ is in the photon
phase, and so the CS term ``Higgses" out the gauge field $a_\mu$.
Or more precisely, for $k = 2$, one gauge flux of $b_\mu$ carries
two gauge charges of $a_\mu$; therefore the photon phase of
$b_\mu$, which is the superfluid phase of gauge flux, breaks the
U(1) gauge invariance of $a_\mu$ to $Z_2$, as was discussed in
Section~\ref{sec:cs}. As also noted in Section~\ref{sec:cs}, this
implies that the GSM of the spinon condensate is $S^3/Z_2$, and
this is the GSM of the spiral spin state. Another way to
understand this phase is that, because the vison number is
attached with the flux number of $a_\mu$ through the mutual CS
term, in the Mott insulator phase of vison the photon of $a_\mu$
is gapped out, while $b_\mu$ is in the photon phase. The global
symmetry $\mathrm{U(1)_b}$ becomes the global U(1) symmetry of the
spinon $z_\alpha$, which according to the PSG in Eq.~(\ref{psga})
corresponds to the physical translation transformation.

\subsubsection{The vison condensate with $s_v < 0$ and $s_z > 0$}
\begin{figure}
\begin{center}
\includegraphics[width=2.5in]{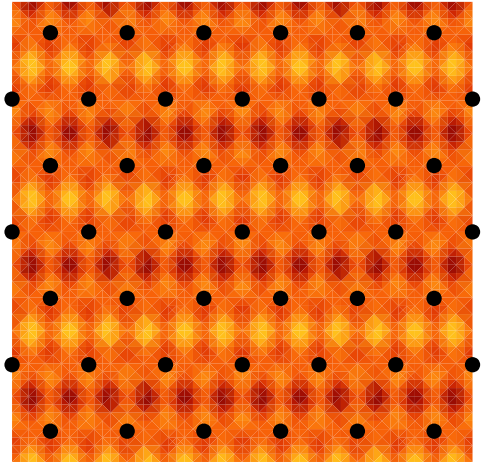}
\caption{Illustration of the $Z_2$ stripe order. This can be
viewed as a bond density wave, with alternating signs on
successive rows of bonds.} \label{figstripe}
\end{center}
\end{figure}

The nature of the phase with spinon gapped and vison condensed
depends on the sign of $g$ in Eq. (\ref{CSAI}). With the spinon
gapped, the U(1) gauge field $b_\mu$ is broken down to $Z_2$ gauge
field. Integrating out the remnant $Z_2$ gauge field, the vison
$v_a$ enjoys a $\mathrm{U(1)\times U(1)}\times Z_2$ symmetry. The
two U(1)s correspond to the global symmetry of two flavors of
visons respectively, and the $Z_2$ symmetry corresponds to the
interchange symmetry between $v_1$ and $v_2$. With $g
> 0$, the vison condensate breaks the $\mathrm{U(1)\times
U(1)}\times Z_2$ symmetry to $\mathrm{U(1)}$ symmetry {\em i.e.\/}
only one flavor of $v_a$ condenses. Let us assume $z_1$ condenses,
and $z_2$ remains gapped. The GSM of the vison condensate is $S^1
\times Z_2$. The $Z_2$ degeneracy is described by the Ising order
parameter $v^\dagger \sigma^z v$, which corresponds to the stripe
order depicted in Fig.~\ref{figstripe}.
The $S^1$ in the GSM is described by the order parameter $v_1^2$,
which corresponds to an incommensurate valence bond density wave
along the $x$ direction with wave vector $(2q_{1x}, \pi)$,
meanwhile breaks the Ising symmetry of interchanging $v_1$ and
$v_2$, or the reflection symmetry $\mathrm{P_y}$. The continuous
symmetry $U(1)$ transformation of $v_1$ which has been broken by
this valence bond density wave is the translation along $\hat{x}$.
Because $v_1$ carries an incommensurate momentum, the full
Lagrangian in (\ref{CSAI}) with all the higher order perturbation
should preserve this global U(1) symmetry, and the GSM $S^1 \times
Z_2$ is not broken down to smaller manifolds. For instance, the
eighth order term $\mathcal{L}_8$ violates the vison numbers of
both $v_1$ and $v_2$, in the phase with only $v_1$ condensed,
$\mathcal{L}_8$ is suppressed.

If $g < 0$, the vison condensate breaks its global symmetry to
$Z_2$ $i.e.$ both $v_1$ and $v_2$ condense with equal stiffness,
the GSM is $S^1 \times S^1$ if there are no more symmetry breaking
terms. The two $S^1$ corresponds to two U(1) order parameters
$v_1v_2$ and $v_1^\ast v_2$ respectively. The U(1) symmetry of
$v_1^\ast v_2$ is preserved by the full Lagrangian, while the U(1)
symmetry of $v_1v_2$ is broken by the eighth order term
$\mathcal{L}_8$ in Eq.~(\ref{l8}). This term breaks the U(1)
symmetry of $v_1v_2$ to $Z_4$ symmetry. Actually, the order
parameter $v_1v_2$ corresponds to the $Z_4$ degeneracy of the four
VBS states, which are smoothly connected to the four fold
degenerate VBS states in the square lattice limit with $J^\prime
\sim 0$. Using the square lattice coordinates, the VBS order
parameters are \beqn V_{\overline{x}} &\sim& v_1v_2 \exp(i \pi/4)
+ v_1^\ast v_2^\ast \exp( - i \pi/4), \cr V_{\overline{y}} &\sim &
v_1v_2 \exp(- i \pi/4) + v_1^\ast v_2^\ast \exp( i \pi/4). \eeqn
The square lattice VBS order is selected when $g_8 > 0$, otherwise
the four fold degenerate plaquette state is favored. On top of
this commensurate VBS order, an incommensurate valence bond
density wave corresponding to $v_1^\ast v_2$ is also present, with
wave vector $(\pi - 2q_{1x}, 0)$.

\subsubsection{The phase with both spinons and visons condensed, $s_z < 0$,
$s_v < 0$}

The most interesting phase is the phase with both spinons and
visons condensed. In this phase, the physical order parameter
should be the U(1) gauge invariant bilinears of $z_\alpha$ and
$v_a$, as discussed in Section~\ref{sec:cs}. The spin order
parameter is the nematic vector $\vec{n}_3 \sim z^\dagger \sigma^a
z$. The VBS pattern, depending on the sign of $g$, is either the
$Z_2$ order parameter $v^\dagger \sigma^z v$, or the
incommensurate valence bond density wave $v^\ast_1 v_2$. Note
however, that the commensurate VBS order parameters
$V_{\overline{x}}$ and $V_{\overline{y}}$ vanish, because they are
not gauge invariant. Another way of understanding the vanishing of
$V_{\overline{x}}$ and $V_{\overline{y}}$ is as follows: when the
spinon $z_\alpha$ is still condensed, the flux of $a_\mu$ is in
the Mott insulator phase; because the flux number of $a_\mu$ is
attached to the vison number through the mutual CS term, any order
parameter violating the vison number conservation should not
condense.

The GSM of this phase is either $S^2_{spin} \times Z_2$ ($g > 0$)
or $S^2_{spin} \times S^1$ ($g < 0$). And these GSM are not lifted
by any higher order term in the Lagrangian Eq. \ref{CSAI}.

Also note that $\mathcal{L}_8$ in Eq.~(\ref{l8}) violates the
gauge symmetry of $b_\mu$. In the condensate of visons, $L_8$
confines the fluxes of $b_\mu$, and hence the spinon excitation
$z_\alpha$ is also confined, which is consistent with the
intuitive understanding of VBS states.

\subsection{Model AII}

The phase diagram of this model is very similar to the previous
subsection, the model AI. The only difference is that we now
replace $v_2$ by $v_2^\ast$. So the spiral spin density wave, the
$Z_2$ spin liquid, and the VBS order are the same as in model AI.
The commensurate VBS order parameter is now represented as \beqn
V_{\overline{x}} &\sim& v_1v_2^\ast \exp(i \pi/4) + v_1^\ast v_2
\exp( - i \pi/4), \cr V_{\overline{y}} &\sim & v_1v_2^\ast \exp(-
i \pi/4) + v_1^\ast v_2 \exp( i \pi/4). \label{vbsAII}\eeqn And
the incommensurate valence bond wave is represented by $v_1v_2$.

\begin{figure}
\begin{center}
\includegraphics[width=3.5in]{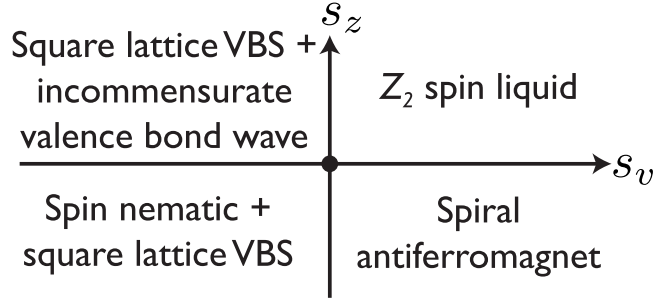}
\caption{Phase diagram of model AII with $g<0$. The phase diagram
of model AII with $g>0$ is the same as that for model AI with $g>0$.}
\label{phaseAIIm}
\end{center}
\end{figure}

In model AII, the VBS pattern with both spinons and visons
condensed is different from model AI. Because in this phase all
the physical order parameters should be U(1) gauge invariant, the
VBS order parameter is either the $Z_2$ symmetry breaking
$v^\dagger \sigma^z v$, or the $Z_4$ symmetry breaking
$V_{\overline{x}}$ and $V_{\overline{y}}$ depending on the sign of
$g$. The GSM is $S^2_{spin} \times Z_2$ ($g
> 0$) or $S^2_{spin} \times Z_4$ ($g < 0$). The phase diagram is shown in
Fig.~\ref{phaseAIIm}.

\subsection{Model AIII}

\begin{figure}
\begin{center}
\includegraphics[width=3.5in]{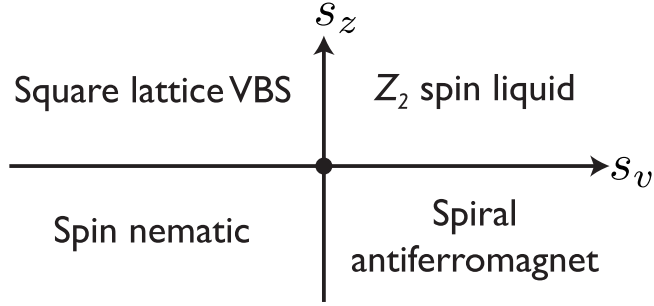}
\caption{Phase diagram of model AIII.}
\label{phaseAIII}
\end{center}
\end{figure}

In this model there is only one complex vison. The $Z_2$ spin
liquid and the spiral spin state are the same as the two previous
models. The vison condensate induces the four fold degenerate VBS
order, with order parameter $V_{\overline{x}} \sim v^2 \exp(i
\pi/4) + h.c.$, $V_{\overline{y}} \sim v^2 \exp(- i \pi/4) +
h.c.$. However, one can no longer write down a U(1) gauge
invariant order parameter in terms of $v$, therefore the phase
with both spinons and visons condensed has only the nematic order
$\vec{n}_3$, and no other lattice symmetry breaking. The phase
diagram is shown in Fig.~\ref{phaseAIII}.

If the parameter $t^\prime / t$ is tuned, the four vison minima
will merge to two vison minima $i.e.$ the two complex visons
become one complex vison. Therefore by tuning $t^\prime / t$,
model AIII can be connected to model AI.

\subsection{Models BI, BII, BIII}
\label{phasediaB}

These models are similar to models AI, AII, AIII. The main
difference is that, in the phase with both spinons and visons
condensed, the spin order is the collinear N\'eel order, in place
of the spin nematic order parameter. The phase diagram for model
BIII was shown in Fig.~\ref{phasesq}.

We now discuss in some detail how the spiral order emerges in model B,
as this is not evident from the underlying spin representation in
Eq.~(\ref{szqb}). If we use only the constraints imposed by the model B
PSG in Eq.~(\ref{psgmodelb}), then the Lagrangian of the spinons
allows an additional inear spatial derivative term:
 In the Lagrangian of spinon of
 \beqn
\mathcal{L}_x \sim \epsilon_{\alpha\beta}z_\alpha\nabla_x z_\beta
+ \mbox{H.c.}  \eeqn The term $\mathcal{L}_x$ violates the
enlarged U(1) gauge invariance of the mutual CS Lagrangian
discussed below Eq.~(\ref{szqb}). The mutual CS term will bind
this term with the monopole operator of $b_\mu$, which creates $2
\pi$ $b_\mu$ gauge flux. We denote this monopole operator by
$\mathcal{M}_b$, then in the U(1) gauge invariant formalism
$\mathcal{L}_x$ reads \beqn \mathcal{L}_x \sim
\mathcal{M}_b\epsilon_{\alpha\beta}z_\alpha\nabla_x z_\beta +
\mbox{H.c.} \label{L_x} \eeqn In the phase with both spinons and
visons condensed, the term $\mathcal{L}_x$ is suppressed, because
of the conservation of the flux of $b_\mu$ which is attached to
the spinon number of $z_\alpha$. However, once the visons are
gapped, the monopoles $\mathcal{M}_b$ condense and $\mathcal{L}_x$
becomes relevant. Due to its linear derivative of $x$,
$\mathcal{L}_x$ will drive the system into an incommensurate
spiral state with wave vector along $\hat{x}$ axis, as has been
described in Ref.~\onlinecite{sr}. The size of the incommensurate
wave vector increases linearly with $\langle \mathcal{M}_b \rangle
\sim \sqrt{\rho}_b$, where $\rho_b$ is the stiffness of the
$b_\mu$ flux condensate, which is proportional to the gap of
vison.

\section{Quantum phase transitions}
\label{sec:qpt}

There are many phase transitions involved in the phase diagrams
discussed in Section~\ref{sec:phase}. We will study them in the
same manner as the previous section.

Before turning to the individual cases, it is useful to discuss an alternative
form of the mutual CS theories, Eq.~(\ref{lcs}). For many of the vison
models, it is possible to \cite{dh,mv} ``undualize'' the vison degrees of freedom: this
leads to an alternative formulation of the theory, now without a CS term.
This new undualized form will be useful for many purposes.

Let us first consider the simplest case of a single complex vison,
as in model III. By the usual boson-vortex duality \cite{dh}, the
dual of $v$ is the monopole operator $\mathcal{M}_b$ introduced
below Eq.~(\ref{L_x}). This monopole operator carries charge $k=2$
under $a_\mu$,  and consequently we can write the theory for the
two-component spinor $z_\alpha$ and the complex ``Higgs'' scalar
$\mathcal{M}_b$: Thus a theory equivalent to Eq.~(\ref{lcs}) for
models AIII and BIII is \beqn \mathcal{L}_M &=& \sum_{\alpha =
1}^2\Big\{ |(\partial_\mu - i a_\mu)z_\alpha|^2 + s_z |z_\alpha|^2
\Bigr\} +
 |(\partial_\mu + 2i  a_\mu)\mathcal{M}_b|^2 - s_v |\mathcal{M}_b|^2
\cr\cr &+&   u_z \left(\sum_{\alpha=1}^2|z_\alpha|\right)^2 + u_M
|\mathcal{M}_b|^4 +v_M |\mathcal{M}_b|^2  \left( \sum_{\alpha=1}^2
|z_\alpha|^2 \right) \cr\cr &+&  \lambda (\mathcal{M}_b
\epsilon_{\alpha\beta} z_\alpha
\partial_x z_\beta + \mbox{H.c.}). \label{CSM}
\eeqn The $\lambda$ term descends from Eq.~(\ref{L_x}), and is
present only for model BIII. A closely related model, for a
similar model, was obtained directly from the Schwinger boson
formulation in Ref.~\onlinecite{sr}. Note that we have
(schematically) changed the sign of the ``mass'' term for
$\mathcal{M}_b$ from that for the vison $v$. This reflects the
dual relation between the fields, and the fact the $v$ is
condensed when $\mathcal{M}_b$ is gapped, and vice versa. We note
that the mapping between the CS theory in Eq.~(\ref{lcs}) and the
non-CS theory in Eq.~(\ref{CSM}) is similar to that described for
supersymmetric gauge theories in Ref.~\onlinecite{m2f}.

A similar (un)duality mapping can applied to visons in model I and
II. This mapping only works for the $g<0$ (``easy-plane'') case of
the theory in Eq.~(\ref{CSAI}). In this case, the vison fields
$v_{1,2}$ and the gauge field $b_\mu$ form an easy-plane CP$^1$
model, and so we can directly use the duality mappings of
Ref.~\onlinecite{mv}. The dual theory is yet another CP$^1$ model,
with fields $m_{1}$ and $m_{2}$ and a gauge field $c_\mu$. Here
$m_{1,2}$ are merons in the vison CP$^1$ model, and the monopole
in the $b_\mu$ field is \cite{mv,senthil} $\mathcal{M}_b \sim
m_{1} m_{2}$. Thus a new form of the theory (\ref{CSAI}) for
models AI, AII, BI, BII with $g<0$ is \beqn \mathcal{L}_m &=&
\sum_{\alpha = 1}^2\Big\{ |(\partial_\mu - i a_\mu)z_\alpha|^2 +
s_z |z_\alpha|^2 \Bigr\} \cr\cr &+&
 |(\partial_\mu + i  a_\mu + i c_\mu )m_{1}|^2 +  |(\partial_\mu + i  a_\mu - i c_\mu )m_{2}|^2
 - s_v \Bigl\{ |m_{1}|^2 + |m_{2}|^2 \Bigr\}
\cr\cr &+&   u_z \left(\sum_{\alpha=1}^2|z_\alpha|\right)^2 + u_m
\Bigl\{ |m_{1}|^2 + |m_{2}|^2 \Bigr\}^2 + g_m |m_1|^2 |m_2|^2
\cr\cr &+& v_m   \Bigl\{ |m_{1}|^2 + |m_{2}|^2 \Bigr\} \left(
\sum_{\alpha=1}^2 |z_\alpha|^2 \right) +    \lambda ( m_1 m_2
\epsilon_{\alpha\beta} z_\alpha
\partial_x z_\beta + \mbox{H.c.}). \label{CSm}
\eeqn
Again the $\lambda$ term is present only for models BI and BII. Also the phase diagrams
can be mapped by keeping in mind the dual relation between the visons $v_{1,2}$
and the merons $m_{1,2}$: the visons are condensed when the merons are gapped,
and vice versa.

Now we will turn to a description of the transitions for the various
models, using the theories in Eq.~(\ref{lcs}), (\ref{CSAI}), (\ref{CSM}) and (\ref{CSm}).

\subsection{Phase transitions in model AI}

\subsubsection{Transition between $Z_2$ spin liquid and spiral spin state}

This transition is known \cite{azaria,css,kim} to be a 3D O(4) transition, and the
mutual CS theory does reproduce this O(4) universality class: in
the $Z_2$ spin liquid, the vison is gapped, therefore the gauge
field $a_\mu$ is ``Higgsed" by gauge field $b_\mu$, and the U(1)
gauge symmetry of $a_\mu$ is broken down to the $Z_2$ gauge
symmetry. The critical point described by spinon $z_\alpha$ enjoys
an enlarged O(4) symmetry.

\subsubsection{Transition between $Z_2$ spin liquid and the VBS state}

The nature of this transition depends on the sign of $g$. When $g
< 0$, the Lagrangian describing this transition is \beqn
\mathcal{L} = \sum_{a = 1}^2 |\partial_\mu v_a|^2 + r|v_a|^2 + u_v
|v_a|^4 + (2u_v + g)|v_1|^2|v_2|^2 + g_8 v_1^4v_2^4. \eeqn This
Lagrangian describes two coupled 3D XY transitions. The coupling
$g_8$ is clearly irrelevant at this 3D XY transition. The scaling
dimension of $u_{12} = 2u_v + g$ is $2/\nu - D < 0$, therefore is
also irrelevant ($\nu$ is the critical exponent defined as $\xi
\sim r^{-\nu}$ at the 3D XY transition, which is greater than
$2/3$). So the transition between the $Z_2$ spin liquid and the
VBS order is two copies of 3D XY transitions when $g < 0$.

When $g > 0$, the transition breaks the $U(1)\times Z_2$ symmetry.
There can be one single first order transition or two separate
transitions, with 3D XY and 3D Ising universality class
respectively. If the triangular lattice is undistorted, $g = 0$,
there is one single transition between the $Z_2$ spin liquid and
the VBS order, which belongs to 3D O(4) universality class.

\subsubsection{Transition between spiral spin state and the nematic+VBS
state}

If $g < 0$ ($g > 0$), this transition is described by a CP$^1$
Lagrangian with easy plane (easy axis) limit: \beqn \mathcal{L} =
\sum_{a = 1}^2 |(\partial_\mu - ib_\mu) v_a|^2 + s_v |v_a|^2 +
\cdots \eeqn The eighth order anisotropy term $\mathcal{L}_8$ is
suppressed at this transition, because the condensate of spinon is
the Mott insulator phase of the flux of $a_\mu$, which guarantees
the conservation of total vison number.

\subsubsection{Transition between VBS order and nematic order}

This is a CP$^1$ transition described by spinon $z_\alpha$ and
U(1) gauge field $a_\mu$. The eighth order anisotropy term of
vison $v_1^4v_2^4 + h.c.$ violates the conservation of the flux of
$a_\mu$ $i.e.$ it corresponds to the instantons in the 2+1d space
time which creates/anihilates gauge fluxes. Because $ k =2$ in
Eq.~ (\ref{CSAI}), one flux of $a_\mu$ carries two visons,
therefore $\mathcal{L}_8$ corresponds to a quadrupole process. If
$g > 0$, only one component of $v_1$ and $v_2$ condenses, the
quadrupole process which involves both $v_1$ and $v_2$ is
suppressed. However, if $g < 0$, the quadrupole process is
present, but expected to be irrelevant at the CP$^{1}$ critical
point \cite{senthil}. We note here recent numerical studies of the
CP$^1$ field theories, which include indications that this
transition is weakly first order
\cite{mv,sandvik,melkokaul,wiese,kuklov,mv2}.

One other issue to notice is that the spinon velocity and the
vison velocity do not have to be equal. Therefore in the CP$^1$
models described above, the velocity of matter fields and the
velocity of gauge fields are essentially different. In the large
$N$ limit, the U(1) gauge field has scaling dimension 1, and the
one-loop self-energy of gauge field leads to the same velocity as
the matter fields. The term with velocity anisotropy has scaling
dimension 4, and hence is irrelevant for large enough $N$.


\subsection{Phase transitions in model AII}

The phase transitions in model AII are similar to model AI. The
only difference is that the eighth order vison term $\mathcal{L}_8
= g_8(v_1 v_2^\ast)^4 + h.c.$ conserves the total flux number of
$a_\mu$, therefore there is no quadrupole process at the
transition between the nematic order and the VBS order.

\subsection{Phase transitions in model AIII}

In model AIII, the transition between the spiral spin order and
the $Z_2$ spin liquid is still O(4), while the transition between
the $Z_2$ spin liquid and the VBS order is a 3D XY transition
(with an irrelevant eighth order anisotropy), because there is
only one flavor of vison.

The transition between the spiral spin order and the nematic spin
order is an inverted 3D XY transition \cite{dh}, or a CP$^0$ model
with one component of complex boson $v$ coupled with the U(1)
gauge field $b_\mu$. The transition between the nematic order and
the VBS order is a CP$^1$ transition with irrelevant quadrupoles.

\subsection{Phase transitions in model BI, BII and BIII}
\label{Bott}

The transitions in the models with spinon B are similar to the
models with spinon A. The major concern is the effect of
$\mathcal{L}_x$ in Eq.~(\ref{L_x}) at the critical points. As
discussed in Section~\ref{phasediaB}, this term is only effective
when the vison is gapped or critical, for instance the transition
between the $Z_2$ spin liquid and the spiral antiferromagnet. The
theory for this transition with spinon A was the O(4) model. With
the term $\mathcal{L}_x$ present in model B, we can redefine the
spinon field using a $x$-dependent O(4) rotation to absorb the
linear $x$ derivative \cite{sr}. The transition is therefore seen
to remain in the O(4) class in model B.

At the transition between the N\'eel order and spiral order, the
field theory is given by a $\mathrm{CP}^{(N-1)}$ model with $N$
flavors of visons ($N = $ 1 or 2) and gauge field $b_\mu$.
$\mathcal{L}_x$ violates the conservation of spinon, and hence
corresponds to a monopole term $\mathcal{M}_b$ of $b_\mu$. For
simplicity, let us consider the model BIII with one vison
component as an example. Here we can analyze the N\'eel-spiral
transition from the theory in Eq.~(\ref{CSM}) by condensing the
monopole operator $\mathcal{M}_b$. We parametrize the spinon $z$
as $z = e^{i\alpha}(e^{i\phi/2}\cos(\theta/2), e^{-
i\phi/2}\sin(\theta/2))^t$, where $\alpha$ is a gauge dependent
phase angle coupled with the gauge field $a_\mu$. Then the
effective Lagrangian can be written as \beqn L =
(\partial_\mu\theta)^2 + (\sin\theta)^2(\partial_\mu\phi)^2 +
\tilde{\lambda}(\nabla_x\theta \mathrm{Re}[\tilde{\mathcal{M}}_b]
+ \nabla_x\phi\sin\theta \mathrm{Im}[\tilde{\mathcal{M}}_b]).
\eeqn where $\tilde{\mathcal{M}}_b$ is the gauge invariant
monopole $\tilde{\mathcal{M}}_b = \mathcal{M}_b e^{2i\alpha}$.
Integrating out the gapless spin-waves $\phi$ and $\theta$, a
singular long range dipole interaction is generated for field
$\tilde{\mathcal{M}}_b$ with momentum dependence
$q_x^2/(q^2+\omega^2)$, which will change the relative scaling
dimension between $x$ and $y$, $\tau$. The effective theory for XY
field $\Psi \sim \tilde{\mathcal{M}}_b$ can be viewed as an
effective $z = 2$ theory with scaling dimension $\Delta[q_x] =
2\Delta[q_y] = 2\Delta[\omega] = 2$: \beqn L_\Phi =
\frac{q_x^2}{q_y^2 + \omega^2}|\Psi|^2 + (q_y^2 +
\omega^2)|\Psi|^2 + g|\Psi|^4 + \cdots \eeqn The upper critical
dimension of this $z = 2$ field theory is $d = 2$, therefore this
transition will be a mean field transition instead of a 3D XY
transition.


For $N_v = 2$ theories in Eq. \ref{CSm}, a similar dipolar term is
generated at the quartic term for $m_i$, a more detailed analysis
is required to determine the fate of this quartic term.

\subsection{Isotropic triangular lattice}
\label{transition:isotropic}

This subsection will briefly comment on the modifications of our results
for the case of the isotropic triangular lattice, with $J'=J$ and full six-fold
rotation symmetry.

There is one more symmetry
that needs to be considered: the $2\pi/3$ rotation. Under this
rotation, the visons of model II transform as \beqn
\mathrm{R}_{2\pi/3} &:& v_1 \rightarrow
\frac{1}{\sqrt{2}}e^{-i\pi/4}v_1 + \frac{1}{\sqrt{2}}v_2,\ \ \ v_2
\rightarrow - \frac{1}{\sqrt{2}}v_1 + \frac{1}{2}e^{\pi i/4}v_2.
\eeqn This PSG transformation is consistent with the enlarged U(1)
gauge symmetry; while in model I, visons will be mixed with its
complex conjugates, therefore on the isotropic triangular lattice
only model II of visons is consistent. Further,
the spinon minima are located at the commensurate wave vectors
$\vec{Q} = (2\pi/3, 0)$ and $- \vec{Q} = -(2\pi/3, 0)$, and therefore
under translation spinons in model A will merely gain a phase
factor, while in model B spinons will be mixed with their complex
conjugates. Therefore on the isotropic triangular lattice, only
model AII is consistent with the enlarged U(1) gauge symmetries.

In the mutual CS theory of model AII on the isotropic lattice, $g
= 0$, in the phase with both spinon and vison condensed, the VSB
order parameter is described by the SU(2) vector $v^\dagger
\sigma^a v$, which corresponds to degenerate stripe orders
$V_{\bar{x}}$, $V_{\bar{y}}$ in Eq. (\ref{vbsAII}), and $V_z =
v^\dagger \sigma^z v$ depicted in Fig. \ref{figstripe}, these
stripe orders are connected to each other through rotation
$\mathrm{R}_{2\pi/3}$. Notice that, on the distorted triangular
lattice, stripe order $V_{\bar{x}}$ has the same symmetry as the
square lattice VBS order.

Because $g = 0$, the global symmetry of vison up to the forth
order term is SU(2) in the mutual CS theory, and O(4) in the
theory with only visons. The GSM of the phase with both spinon and
vison condensed is $S^2 \times S^2$ as far as the forth order
terms are considered. Therefore the transition between the $Z_2$
spin liquid and the VBS order is a 3D O(4) transition
\cite{sondhi}, and the transition between the nematic/VBS order to
the spiral order in phase diagram Fig. \ref{phaseAIIm} is a CP$^1$
transition. The PSG of visons allow a sixth order anisotropy term
on the isotropic triangular lattice \cite{sondhi}: \beqn L_6 = g_6
(v_1v_2^5 + v_2v_1^5+ H.c. ) \eeqn This term corresponds to the
triple monopole process in the dual picture, which
annihilates/creates three fluxes of gauge field $a_\mu$. This
triple monopole is expected to be relevant when gauge field
$a_\mu$ is gapless, which will likely drive the transition between
the nematic/VBS and the VBS phase to a first order transition.

\subsection{Multicritical point, $s_z = s_v = 0$}
\label{sec:multi}

We now study the multicritical point with both spinons and visons
gapless: this is the point M in Fig.~\ref{phasecs}.

The most convenient way of studying M is likely via the non-CS
theories in Eqs.~(\ref{CSM}) and (\ref{CSm}), although these do
not apply for the $g > 0$ cases. This formulation should be
amenable to direct numerical study.

For analytic results, the only available tool is the $1/N$
expansion, for this we may as well work with the original CS
theory in Eq.~(\ref{lcs}). This expansion relies on the assumption
that in Eq. (\ref{CSAI}) $N \sim N_v \sim k$ is large. The
$\lambda$ term in Eq.~\ref{CSM} and ~\ref{CSm} generalizes to
terms with $k$ powers of $z_\alpha$, and these are surely
irrelevant for large $k$. So we ignore the influence of
$\mathcal{L}_x$ for models B in the $1/N$ expansion.

A systematic $1/N$ expansion for the CP$^{(N-1)}$ model has been
calculated previously \cite{ir,ks}: the $1/N$ correction comes
from the one-loop propagator of both the Lagrange multiplier
$\lambda$ and gauge field $a_\mu$: \beqn \mathcal{L} =
\frac{1}{g}|(\partial_{\mu} - ia_\mu)z|^2 + i\lambda(|z|^2 - 1)+
\cdots \eeqn The one loop propagator of $\lambda$ and $a_\mu$ are:
\beqn D_{\mu\nu} &=& \frac{1}{\Pi_A} \left(\delta_{\mu\nu} - \zeta
\frac{q_\mu q_\nu}{q^2}\right), \ \ \Pi_A = \frac{N p}{16}, \cr
D_\lambda &=& \frac{1}{\Pi_\lambda}, \ \ \Pi_\lambda = \frac{N p
}{8}. \eeqn For instance, we can calculate the anomalous dimension
$\eta_N$ of gauge invariant operator $z^\dagger T^a z$ defined as
$\Delta[z^\dagger T^a z] = (D - 2 + \eta_N)/2$, and $T^a$ is one
of the generators of SU(N) algebra; note that this operator is
the magnetic order parameter for model B, and the spin nematic order parameter
for model A. In the CP$^{(N - 1)}$ model,
the anomalous dimension $\eta_N$ was calculated in detail in
Reference \onlinecite{ks}, and the result is \beqn \eta_N = 1 +
\frac{32}{3\pi^2 N} - \frac{128}{3\pi^2 N}. \eeqn The second term
on the r.h.s. of the equation above comes from the Lagrange
multiplier, while the third term comes from the gauge field.

After including the vison multiplet $v_a$ and mutual CS term, the
$\lambda$ propagator is unaffected, while the gauge field
propagator is modified: \beqn D_{\mu\nu} &=&
\frac{1}{\tilde{\Pi}_A} \left(\delta_{\mu\nu} - \zeta \frac{q_\mu
q_\nu}{q^2}\right), \ \ \widetilde{\Pi}_A = \frac{N p}{16}\left(1
+ \frac{64k^2}{\pi^2 N^2}\right). \eeqn All the calculations can
be carried out straightforwardly by replacing $\Pi_A$ in
Ref.~\onlinecite{ks} by $\widetilde{\Pi}_A$. When $k \sim N$, the
correction from the vison and mutual CS theory to the anomalous
dimension $\eta_N$ is at the order of $1/N$ : \beqn \eta_N = 1 +
\frac{32}{3\pi^2 N} - \frac{128}{3\pi^2 N}\times \frac{1}{1 +
{64k^2}/({\pi^2 N^2})}. \label{etaN} \eeqn All the other critical
exponents can be calculated in a similar way.

\section{Beyond U(1) Chern-Simons theories}
\label{sec:beyond}

In the previous sections, the global phase diagrams and nature of
phase transitions were studied with the mutual CS theory. In this
section we will try to look at these phases and transitions in a
more intuitive and pictorial way. Let us start with the case with
undistorted triangular lattice, which has a ground state with
spiral antiferromagnetic order. This spiral order is described in
terms of the $z$ by Eq.~(\ref{szq}).

\begin{figure}
\begin{center}
\includegraphics[width=3.5in]{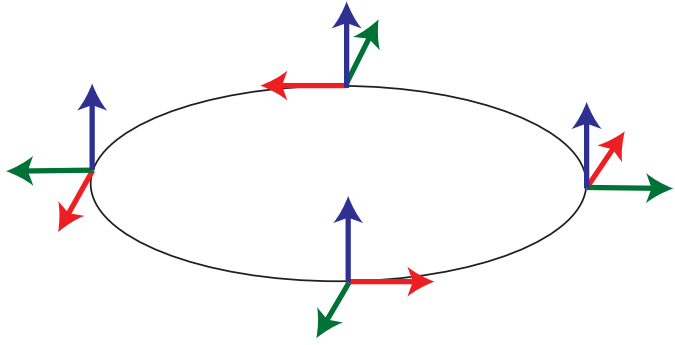}
\caption{Schematic of the orientation of the orthogonal vectors
$\vec{n}_1$, $\vec{n}_2$ and $\vec{n}_3$ around a vortex in the
SO(3) GSM of the spiral antiferromagnet. One of the vectors has a
constant orientation, while the other two precess by an angle of
$2 \pi$.} \label{vortex}
\end{center}
\end{figure}

Intuitively, to destroy magnetic order, the most straightforward
way is to proliferate the topological defects in this magnetic
order. The fundamental group of the GSM of spiral order is
$\pi_1[\mathrm{SO(3)}] = Z_2$, which supports a topologically
stable half vortex if rewritten in terms of $z$. These half
vortices become full vortices of vectors $\vec{n}_i$, $i = 1,2,3$
in Eq.~(\ref{szq}). These vortices can be most easily understood
as an ordinary vortices of two of the three vectors $\vec{n}_i$,
while keeping the third vector uniform in the whole 2d plane: see
Fig.~\ref{vortex}.
The GSM of this state has isometry group SO(4). If the system has
the enlarged O(4) symmetry at the microscopic level, all the
vortices with different uniform vector (UV) will have the same
energy. However, the underlying symmetry is only $\mathrm{SU(2)}
\times \mathrm{PSG}$, so the energy of the vortices depends on the
UV. The lattice symmetry guarantees that the vortices with
different UVs have the same energy as long as the UVs can be
transformed to each other through lattice symmetry
transformations. There are in total three groups of vortices:
\beqn 1, \ \ \mathrm{UV} &:& \ \ \vec{n}_3, \cr\cr 2, \ \
\mathrm{UV} &:& \ \ \vec{n}_1, \ \ - \frac{1}{2} \vec{n}_1 \pm
\frac{\sqrt{3}}{2} \vec{n}_2, \cr\cr 3, \ \ \mathrm{UV} &:& \ \
\vec{n}_2, \ \ \pm \frac{\sqrt{3}}{2} \vec{n}_1 - \frac{1}{2}
\vec{n}_2. \label{visons}\eeqn All the flavors of vortices in each
group have the same energy, while there is no symmetry to protect
the degeneracy between different groups. For instance, UV
$\vec{n}_3$ can never be transformed in to $\vec{n}_1$ because of
their opposite behavior under time reversal transformation. The
first group of vortex has only one flavor, and if we only
proliferate this vortex flavor, the spin orders of $\vec{n}_1$ and
$\vec{n}_2$ are destroyed, while the nematic order $\vec{n}_3$ is
preserved. This leads to a state with GSM $S^2$, and it is the
situation described by the mutual CS theory AI, AII and AIII. The
second and third groups have more than one flavor of vortices,
with different flavors of vortices connected to each other through
lattice symmetry transformations. If all the flavors of vortices
in group 2 or 3 condense, the magnetic order is completely
destroyed, and we expect these vortices to drive a direct
transition from the $\sqrt{3}\times \sqrt{3}$ antiferromagnetic
spiral order to VBS order. The nature of this transition requires
further study, and it cannot be naturally described by our mutual
CS theory. In our theory, there are always two tuning parameters
($s_z$ and $s_v$), and it would require some fine-tuning to induce
a direct transition. A theory based on a mutual $Z_2$ CS formalism
has been proposed by another group \cite{senthildirect}
 to describe this direct transition.
\begin{figure}
\begin{center}
\includegraphics[width=5.5in]{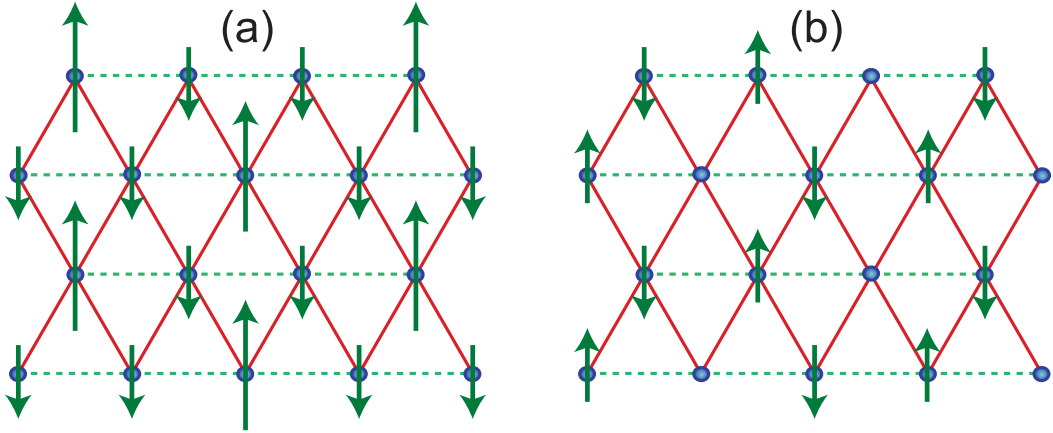}
\caption{The remnant spin order patten after proliferation of one
single flavor of group 2 and 3 vortices in Eq.~(\ref{visons}).
({\em a\/}) Proliferation of the first flavor in group 2, the spin
pattern is up-down-down, and the GSM is $S^2 \times Z_3$. Notice
that the moment of the up spin site is twice as much as the down
spin site, so the total magnetization is zero. ({\em b\/})
Proliferation of first flavor in group 3, the spin pattern is
up-down-zero with GSM $S^2 \times Z_3$. } \label{pattern}
\end{center}
\end{figure}

One can also condense one flavor of the second or the third group
of vortices, which can be realized in the situation with strong
repulsions between different flavors of vortices in one group. If
the vortex with UV $\vec{n}_1$ is condensed, the vectors
$\vec{n}_2$ and $\vec{n}_3$ are disordered, and the remnant
magnetic order is the up-down-down state in Fig.~\ref{pattern}$a$
with zero total magnetization, and the GSM of this up-down-down
state is $S^2 \times Z_3$. The $S^2$ corresponds to the direction
of $\vec{n}_1$, and the $Z_3$ corresponds to the choice of
condensing the three flavors of vortices in the second group of
Eq.~(\ref{visons}), which are connected to each other via
translation along the $x$ axis. If vortex with UV $\vec{n}_2$ is
condensed, the spin pattern becomes the up-down-zero state in
Fig.~\ref{pattern}$b$, also with GSM $S^2 \times Z_3$. The
transition driven by the condensation of vortices with UV
$\vec{n}_1$ and $\vec{n}_2$ can no longer be described by the U(1)
mutual CS theory, because in the phase with both spinons and
visons condensed, the physical order parameter of the remnant spin
order should be U(1) gauge invariant: thus now one needs a spinon
$z_\alpha$ such that the UV $n_1 = z^\dagger \sigma^a z$. However,
under translation so-defined spinon $z_\alpha$ becomes a linear
combination between $z_\alpha$ and
$\epsilon_{\alpha\beta}z^\ast_\beta$, which violates the U(1)
gauge symmetry.

To consistently describe the transition driven by proliferation of
vortices with UV $\vec{n}_1$ and $\vec{n}_2$, a theory based on
mutual $Z_2$ CS theory may be applicable, similar to the $Z_2$ CS
formalism proposed for cuprates \cite{sf} where the
spinons and half-vortices of the superconducting phase are coupled
together through mutual $Z_2$ CS fields. On the distorted
triangular lattice antiferromagnets examined in this work, the
mutual $Z_2$ gauge field also imposes the correct semionic
statistics between the spinon and vison. However the $Z_2$ gauge
field can only be conveniently formulated on the lattice,
therefore we are unable to comment on all the universal properties
of the transitions described by the mutual $Z_2$ CS theory.

If the triangular lattice is distorted, the spin spiral state
becomes incommensurate, and the vector $\vec{n}_1$ and $\vec{n}_2$
can be transformed to each other via lattice translation.
Therefore there are only two groups of vortices: \beqn 1, \ \
\mathrm{UV} &:& \ \ \vec{n}_3, \cr\cr 2, \ \ \mathrm{UV} &:& \ \
\vec{n}_1, \ \ \mathrm{rotation \ of} \ \vec{n}_1 \
\mathrm{around} \ \vec{n}_3. \label{visons1}\eeqn The second group
of vortices have an infinite number of flavors, and if one of
these flavors proliferates, then the spin state becomes a
collinear incommensurate spin density wave, which has GSM $S^2
\times S^1$. The $S^2$ corresponds to the remnant spin collinear
order, while the $S^1$ corresponds to translation along $\hat{x}$
of the incommensurate wave vector.

\section{Conclusions and Experimental implications}
\label{sec:conc}

This paper has described examples of a general approach to
describing the phases and quantum phase transitions of $S=1/2$
antiferromagnets in two dimensions. Our examples were limited to
models on the lattice of Fig.~\ref{lattice}, because of its
experimental importance.  However, we expect that similar analysis
should be useful {\em e.g.} on the kagome lattice. Our main
results are summarized in phase diagrams, like that in
Fig.~\ref{phasesq}. The same phase diagram was obtained earlier
\cite{sr} in a more direct microscopic mean field theory, but the
nature of the phase transitions and possible multicritical point
was left open. Our present, `dual'  approach also immediately
yields the required critical field theories.

One of the transitions in our phase diagram in Fig.~\ref{phasesq}
is the N\'eel-VBS transition, which is described by a CP$^1$ field
theory. This transition has been the focus of much recent
numerical work \cite{mv,sandvik,melkokaul,wiese,kuklov,mv2}. The
numeric results on the $S=1/2$ quantum antiferromagnet strongly
support its effective description in terms of the CP$^1$ field
theory. However, some results \cite{wiese,kuklov} on system sizes
larger than $50\times 50$ indicate that the transition in the
CP$^1$ model may well be weakly first-order. The multicritical
point M has additional flavors of matter fields, and these make it
less likely that M is first order. Consequently it would be useful
to study M numerically using the theories in Eqs.~(\ref{CSM}) and
(\ref{CSm}): this will help in describing M and the phase diagram
in its immediate vicinity. This study can be done with or without
the $\lambda$ coupling in Eqs.~(\ref{CSM}) and (\ref{CSm}).

As mentioned in Section~\ref{sec:intro}, a series of measurements on the
distorted triangular lattice materials, X[Pd(dmit)$_2$]$_2$,  with
different anisotropic coupling $J^\prime / J$ reveal a possible
direct transition between the N\'eel order and the VBS state, and
one particular material with X = EtMe$_3$Sb was suggested to be
close to the quantum critical point. In our mutual CS theory, as
well as the previously proposed deconfined critical point
\cite{senthil}, this transition is described by the CP$^1$ model
with irrelevant quadrupole operators. The quantum critical
behaviors at finite temperature can be measured in material
EtMe$_3$Sb[Pd(dmit)$_2$]$_2$. For instance, the NMR relaxation
rate $1/T_1$ scales as $1/T_1 \sim T^{\eta_N}$, where $\eta_N$ is the
anomalous dimension of the N\'eel order parameter at the CP$^1$
fixed point. Even if the transition in the CP$^1$ model is ultimately
weakly first order, the critical regime could be observable at intermediate $T$,
and compared to numerical estimates \cite{sandvik,mv2,melkokaul} of $\eta_N$.

We conclude by commenting on the recent interpretation by
Ref.~\onlinecite{z2} of experimental observations on the spin
liquid compound $\kappa$-(ET)$_2$Cu$_2$(CN)$_3$.
Ref.~\onlinecite{z2} suggested that the very low $T$ nuclear magnetic resonance was
controlled by the O(4) criticality between the spiral and $Z_2$ spin
liquid states, while the intermediate temperature nuclear
magnetic resonance (NMR) could be modeled by a multicritical point
where both the spinons and visons are gapless.
Crucial to this interpretation was the requirement that the anomalous dimension
of the magnetic order parameter, $\eta_N$, was smaller at the latter multicritical point
than at the low $T$ O(4) critical point.

Here, we have
provided examples of such multicritical points, such as the point
M in Fig.~\ref{phasesq}. For spinons in model B, the magnetic order parameter
is $z^\dagger T^a z$, and its anomalous dimension
was computed in Section~\ref{sec:multi}, where we found the
result in Eq.~(\ref{etaN}).  An important feature of this result is that the U(1)
gauge fluctuations reduce the value of $\eta_N$ from that obtained
in the theory without the U(1) gauge field; the latter describes
the O(4) transition between the spiral antiferromagnet and the $Z_2$
spin liquid. Thus model B spinons do fulfill the requirements for
the experimental interpretation stated in Ref.~\onlinecite{z2}.

On the other hand, in spinon model A, the order parameter $z^\dagger
T^a z$ represents the nematic order parameter, while the other
gauge invariant spinon-monopole composite $z^t \sigma^y \sigma^a z
\mathcal{M}_b$ represents the spiral order parameter. In the case
with large $N_v$, the scaling dimension of $\mathcal{M}_b$ is
expected to scale linearly with $N_v$, and hence to systematically
calculate the scaling dimension of $z^t \sigma^y \sigma^a z
\mathcal{M}_b$, we need an analytical technique beyond the $1/N$
expansion.

In conclusion, we have proposed a unified theory in the mutual CS
formalism to describe all the magnetic phases observed in a series
of organic compounds, and discussed the phase transitions between
these phases. Experimental implications for  organic compounds
in the X[Pd(dmit)$_2$]$_2$ series and the $\kappa$-(ET)$_2$Z series
were also discussed.

\acknowledgments We thank L.~Balents, M.~P.~A.~Fisher, R.~Kato,
I.~Klebanov, P.~A.~Lee, Y.~Qi, T.~Senthil, Y.~Shimizu, M.~Yamashita, and X.~Yin for useful
discussions. This research was supported by the NSF under grant
DMR-0757145, by the FQXi foundation.

\end{document}